\documentclass[twocolumn]{aastex631}

\usepackage{amsmath}
\usepackage{colortbl}
\usepackage{multirow}
\usepackage{subfigure}
\usepackage{graphicx}
\usepackage{xspace}
\usepackage{color}
\usepackage{url}
\usepackage{float}
\usepackage{ulem}
\usepackage{soul}
\usepackage{bm}
\usepackage{longtable}
\usepackage{threeparttable}
\usepackage{booktabs}
\usepackage{hyperref}

\begin{document}

\shorttitle{Highly polarized orthogonal emission of Vela X-1}
\shortauthors{Wu et al.}

\title{Highly polarized intrinsic emission and its orthogonal counterpart in Vela X-1}

\author[0009-0007-5635-0340]{WanYun Wu}
\affiliation{Guangxi Key Laboratory for Relativistic Astrophysics, School of Physical Science and Technology, Guangxi University, Nanning 530004, China}

\author[0000-0002-0105-5826]{Fei Xie}
\correspondingauthor{Fei Xie}
\email{xief@gxu.edu.cn}
\affiliation{Guangxi Key Laboratory for Relativistic Astrophysics, School of Physical Science and Technology, Guangxi University, Nanning 530004, China}
\affiliation{INAF Istituto di Astrofisica e Planetologia Spaziali, Via del Fosso del Cavaliere 100, 00133 Roma, Italy}

\author[0000-0001-9599-7285]{Long Ji} 

\affiliation{School of Physics and Astronomy, Sun Yat-Sen University, Zhuhai, 519082, China}

\author[0000-0002-3776-4536]{MingYu Ge} 

\affiliation{Key Laboratory of Particle Astrophysics, Institute of High Energy Physics, Chinese Academy of Sciences, Beijing 100049, China}

\affiliation{University of Chinese Academy of Sciences, Chinese Academy of Sciences, Beijing 100049, China}

\author[0000-0001-8916-4156]{Fabio La Monaca} 

\affiliation{INAF Istituto di Astrofisica e Planetologia Spaziali, Via del Fosso del Cavaliere 100, 00133 Roma, Italy}

\affiliation{Dipartimento di Fisica, Universit$\grave{a}$ degli Studi di Roma “Tor Vergata”, Via della Ricerca Scientifica 1, 00133 Roma, Italy}

%%%%%%%%%%%%%%%%%%%%%%%%%%%%%%%%%%%%%%%%%%%%%%%%%%%%%%%%%%%%%%%%%%%%
\begin{abstract}

Vela X-1 is one of the most archetypal wind-fed X-ray pulsars (XRPs), and the emergence of its orthogonal polarization states reveals distinctive polarimetric properties. Using data from Imaging X-ray Polarimetry Explorer (IXPE) observations of Vela X-1, we perform a polarization analysis of Vela X-1 using a triple power-law spectral model absorbed by varying column densities, successfully isolating two physically distinct orthogonal polarized components. The first polarized component corresponds to emission from the accretion mound surface that is not obscured by the wind clumps, with its polarization degree (PD) exceeding 30\%. In specific phase intervals, the PD reaches \(50.9 \pm 10.7\%\). This marks the first detection of such highly polarized neutron star emission in an XRP. The second polarized component likely originates from complex physical processes within or near the accretion mound, with its PD showing a potential negative correlation with column density. Furthermore, by rotating the predicted polarization angle (PA) of the first polarized component by 90$^\circ$, we successfully achieve separate fitting and simultaneous fitting of the two orthogonal polarization states using the rotating vector model (RVM).

\end{abstract}

\keywords{binary star; polarization}
%%%%%%%%%%%%%%%%%%%%%%%%%%%%%%%%%%%%%%%%%%%%%%%%%%%%%%%%%%%%%%%%%%%%

\section{Introduction}  \label{sec:intro}

Accreting X-ray pulsars (XRPs) are neutron stars (NS) in binary systems that accrete material from a companion star via an accretion disk or stellar wind. These systems are characterized by strong surface magnetic fields ($10^{12}\sim 10^{13}$G) and exhibit a wealth of astrophysical phenomena governed by complex processes, including accretion flow transitions, plasma-magnetosphere interactions, and the variation of accretion geometry. X-ray polarization has emerged as a powerful probe of these phenomena (see recent review by \cite{Weng2024XRayVO}). The radiation from XRPs is predicted to be highly polarized (e.g., \cite{{Mszros1988AstrophysicalIA}, {Caiazzo2020PolarizationOA}}) due to the stark opacity contrast between the ordinary (O-mode) and extraordinary (X-mode) photon propagation modes \citep{Lai2003a}. These modes arise from plasma effects and vacuum birefringence in ultra-strong magnetic fields \citep{{Gnedin1974},{Gnedin1978}}.

 Recently, the Imaging X-ray Polarimetry Explorer (IXPE) \citep{{Soffitta2021TheIO},{Weisskopf2022ImagingXP}}, launched in December 2021, has provided a dedicated instrument for studying the X-ray polarization properties of XRPs. IXPE has observed some XRPs, such as Her X-1 \citep{{Doroshenko2022DeterminationOX}, {Garg2023}, {Heyl2024ComplexRD}, {Zhao2024PolarizationPO}}, Cen X-3 \citep{Tsygankov2022TheXP}, 4U 1626–67 \citep{Marshall2022ObservationsO4}, Vela X-1 \citep{{Forsblom2023IXPEOO}, {Forsblom2025RevealingTO}}, LS V+1477 \citep{Doroshenko2023ComplexVI}, GX 301$-$2 \citep{Suleimanov2023XrayPO}, GRO J1008–57 \citep{Tsygankov2023XrayPG}, X Persei \citep{Mushtukov2023XrayPO}, EXO 2030+375 \citep{Malacaria2023APO}, Swift J0243.6+6124 \citep{Poutanen2024StudyingGO} and SMC X-1 \citep{Forsblom2024ProbingTP}. These systems exhibit a polarization degree (PD) significantly lower than traditional theoretical predictions, which is attributed to vacuum resonance in atmospheric temperature gradients or variations in viewing angle \citep{Doroshenko2022DeterminationOX}. Among them, several wind-affected sources, such as LS V +44 17 and Vela X-1, show significant swings in polarization angle (PA), commonly attributed to the presence of dual-component polarization signatures.

As a typical wind-accretion source, Vela X-1 (4U 0900-40) is a high-mass X-ray binary (HMXB) located approximately at 2 kpc \citep{Kretschmar2021RevisitingTA} from Earth. It consists of a NS with a spin period of 283 seconds \citep{1976ApJ...206L..99M} and its companion, the B0.5 Ia supergiant HD 77581 \citep{1972ApJ...175L..19H}. The system has an orbital period of 8.964 days \citep{faucris.216777486}, with the NS orbiting at a distance of about 1.7 times the radius of the supergiant \citep{Quaintrell2003TheMO}. Due to the nearly circular orbit (e $\sim$ 0.0898) of the slowly rotating NS around its blue supergiant companion, the line-driven stellar wind provides a continuous source of material for accretion, with the companion's mass-loss rate estimated at \(10^{-6} M_{\odot} \text{ yr}^{-1}\) \citep{GmenezGarca2016MeasuringTS}. Although Vela X-1 exhibits a moderate intrinsic X-ray luminosity of approximately \(4 \times 10^{36} \text{ erg s}^{-1}\) \citep{Sato1986XrayPO}, its relative proximity makes it one of the brightest persistent X-ray sources in the sky. Consequently, it is one of the most extensively studied wind-accreting pulsars and is often regarded as the archetype of wind-fed HMXBs, where stellar wind accretion serves as the primary mechanism for mass transfer. 

During the two observations by IXPE in 2022, Vela X-1 not only exhibits time-averaged low PD (\(<\) 10\%) but also shows a 90$^\circ$ PA swing between the high-energy and low-energy bands \citep{Forsblom2023IXPEOO}. In concurrent studies, \cite{Forsblom2025RevealingTO} proposed that the energy dependence of PD and PA can be modeled by the combination of a power-law component and a thermal bremsstrahlung component, which have different polarization properties with PA differing by 90$^\circ$. At approximately 3.4 keV, the polarized flux contributions from these two components are equal, resulting in a net PD of zero. In addition, they proposed that the time-averaged low polarization and the presence of two orthogonal polarization states originate from the vacuum resonance under the atmospheric density gradient \citep{Doroshenko2022DeterminationOX}, or the mixture of different emissions near the accretion mound.

In this paper, we mainly use a new composite spectral model to perform various polarization analyses of Vela X-1 with two IXPE observations. The structure of this paper is as follows. Section \ref{sec:Data} describes the observations and data reduction procedures. The results are presented in Section \ref{sec:Results} and discussed in Section \ref{sec:discussion}. Finally, we give the summary in Section \ref{sec:summary}.

%%%%%%%%%%%%%%%%%%%%%%%%%%%%%%%%%%%%%%%%%%%%%%%%
%%%%%Observation total duration (seconds!!)
\section{Observation and Data Reduction}
\label{sec:Data}
 
IXPE is a mission jointly initiated by NASA and the Italian Space Agency, launched on December 9, 2021. The detector consists of three identical grazing incidence telescopes and detector modules, each telescope includes an X-ray mirror assembly and a polarization-sensitive detector unit equipped with a Gas-Pixel detector, with an operational energy range of 2--8 keV  \citep{{Soffitta2021TheIO},2022AJ....164..103D,{Weisskopf2022ImagingXP}}.  It is capable of performing imaging polarization measurements with a time resolution of better than 10 microseconds within the detector field of view of 12\arcmin.9 $\times$ 12\arcmin.9. A detailed description of the observatory and its performance can be found in \cite{Weisskopf2022ImagingXP} and reference therein.

Vela X-1 was observed twice by IXPE, with the observation periods being from April 15 to 21, 2022 (OBSID: 01002501) and from November 30 to December 6, 2022 (OBSID: 02005801). We downloaded the Level 2 dataset from HEASARC, and performed a solar system center correction on the photon arrival times, using the standard HEASoft (v.6.30.1) \textit{barycorr}. To correct for orbital motion in the binary system, we used the orbital parameters for Vela X-1 provided by the Fermi Gamma-ray Burst Monitor. The pixel position offset correction and energy calibration were applied. In addition, we remove time intervals where the IXPE count rate in the background annular region was higher than the mean background value plus three times the root mean square (RMS) of this count rate, as these intervals are often considered to be affected by solar activities. After these processes, the source photons were extracted from a circular region with a radius of 70\arcsec\ centered on the source, and no background subtraction was applied, as suggested in \cite{DiMarco2023HandlingTB}.

We process Level 2 data using version 30.2.32 of the \textit{ixpeobssim} software package \citep{Baldini2022ixpeobssimAS}, which incorporates the calibration database (v12) released on 17 November 2022. To derive polarization properties, we applied the PCUBE algorithm in \textit{ixpeobssim} to generate polarization cubes, and utilized the PHA1, PHA1Q, and PHA1U algorithms to extract Stokes unweighted \textit{I}, \textit{Q}, and \textit{U} spectra \citep{DiMarco2022weights}, respectively. Then, the \textit{I} spectrum was binned to ensure that each bin contained at least 30 counts, and then the same energy binning was applied to the Stokes \textit{Q} and \textit{U} spectra. The energy spectra were fitted in the XSPEC package \citep{Arnaud1996}. In this paper, unless stated otherwise, uncertainties are reported at the 68\% confidence level.

\begin{figure}[h]
\centering
\includegraphics[width=0.48\textwidth]{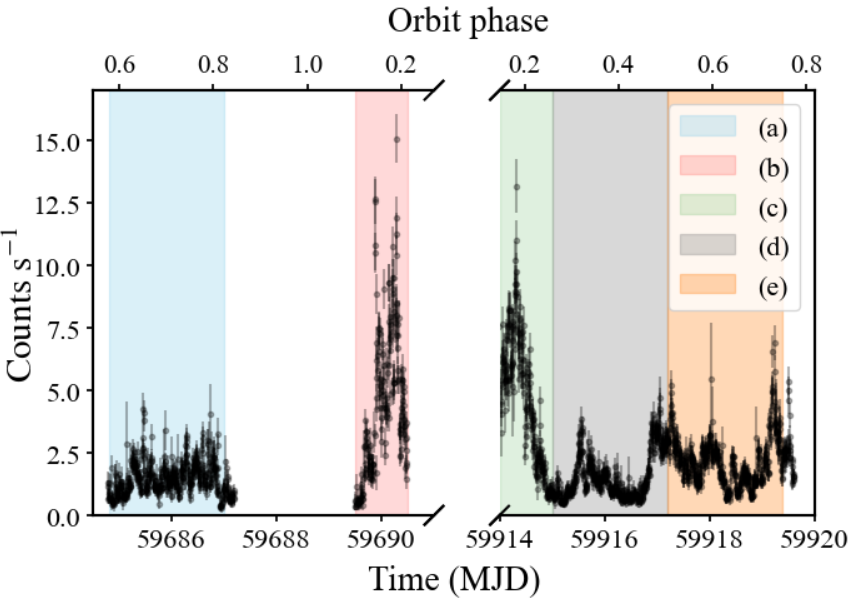}
\caption{Light curves from the two IXPE observations of Vela X-1. Time intervals (a)–(e) are used for time-resolved polarization analysis, respectively corresponding to the MJD time intervals: (a) 59684.8–-59687, (b) 59689.5--59690.5, (c) 59914–-59915, (d) 59915–-59917.2, and (e) 59917.2--59919.4.}
\label{lc}
\end{figure}

\begin{figure}[t]  % [t]
\centering
\includegraphics[width=0.47\textwidth]{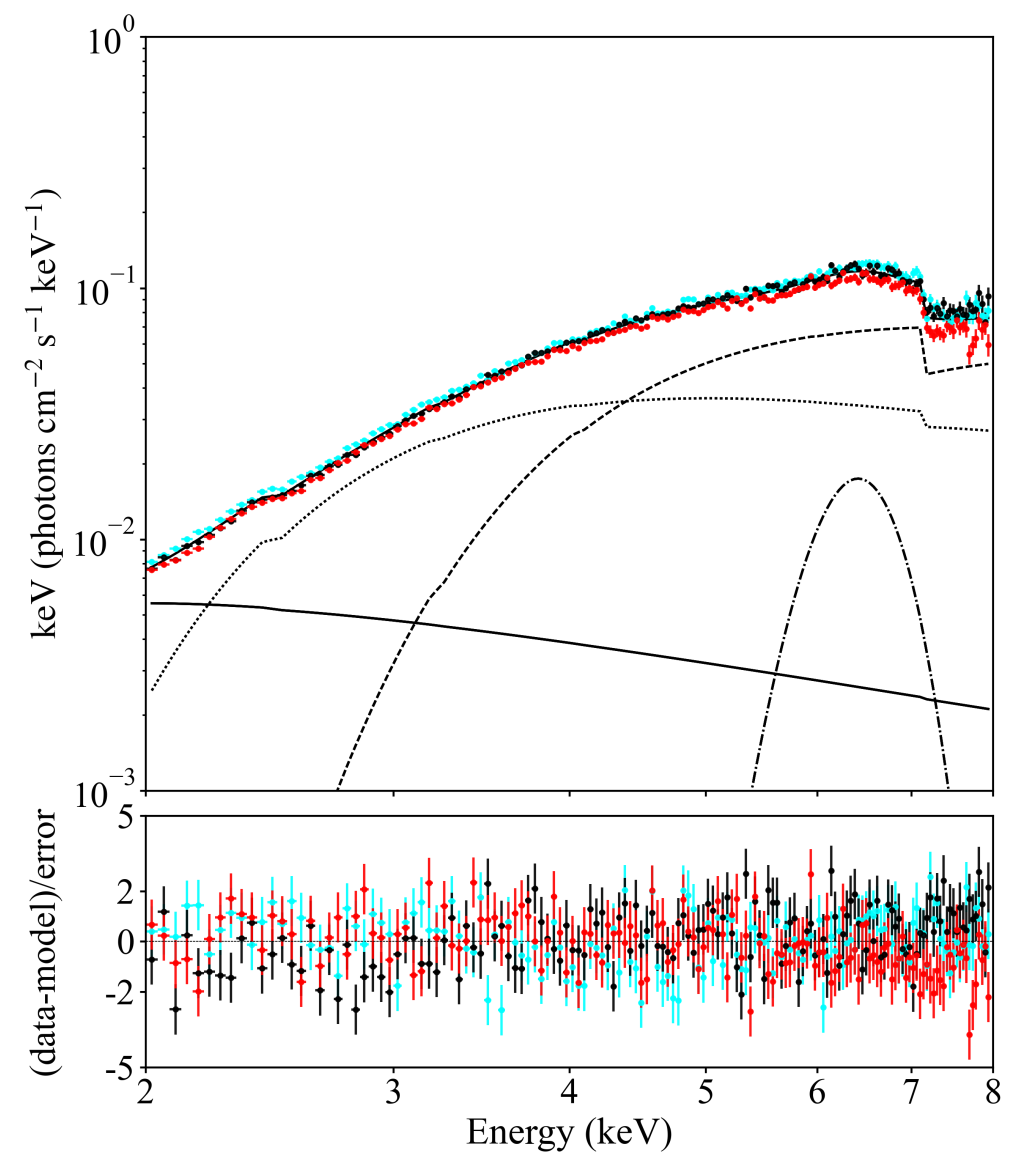}
\caption{Top panel: the spectra of the combined dataset of two IXPE observations. The red, black and cyan colors represent the data from three DUs, respectively. The black solid line, dotted line, dashed line and dot–dashed line represent Component 1, Component 2, Component 3 and the Gaussian component of our fitting model, respectively. The fitting residuals are shown in the bottom panel.}
\label{spec}
\end{figure}

\section{Results}  \label{sec:Results}

The light curves of the two Vela X-1 observations obtained by IXPE in the 2--8 keV energy band are shown in Figure \ref{lc}. We excluded the eclipse data (MJD 59687.3–59689.5) from the first observation in all subsequent analyses. The pulse phases were calculated using the ephemerides listed in Table 1 of \cite{Forsblom2023IXPEOO}.

Subsequently, we present in this section the results of the time-averaged polarization analysis, the time-resolved polarization analysis, and the phase-resolved polarization analysis.

\subsection{The triple power-law model and its polarization} \label{Spectral polarimetric analysis}

The emission from wind-accreting X-ray pulsar systems strongly depends on the stellar wind and typically exhibits significant intrinsic flux variations. These variations cannot always be straightforwardly distinguished, particularly in the soft X-ray band \citep{Diez2023ObservingTO}. Therefore, we adopt a phenomenological model proposed by \cite{MartinezNunez2014TheAE}, specifically a sum of three power-law components, to represent the combination of Vela X-1's complex emission components. Each power-law has different normalizations but shares the same photon index, while being absorbed by distinct absorption components. This model was originally developed for the XMM-Newton spectrum of Vela X-1 in the 0.5--10 keV energy range, and was also used by \cite{Kretschmar2021RevisitingTA} to investigate the impact of variable absorption on Vela X-1's overall spectral shape.

Specifically, we implement the following model:

\begin{equation}
\begin{split}
    & const \times tbabs\times[polconst \times powerlaw \\
    & + polconst\times (tbabs \times powerla w + tbabs \times powerlaw) \\
    & + polconst\times gauss],
\end{split}
\end{equation}

\begin{table*}[t]
\centering
    \begin{tabular}{cc|ccc}
        \toprule
        \textbf{Component} & \textbf{Parameter} & \( \mathrm{\textbf{01002501}} \) & \( \mathrm{\textbf{02005801}} \) & \( \mathrm{\textbf{Total}} \) \\
        \midrule
             tbabs&\( N_\mathrm{{H,1}} \) \( (10^\mathrm{22} \mathrm{cm^{-2}}) \) & [0.75] & [0.75] & [0.75] \\
             &\( N_\mathrm{{H,2}} \) \( (10^\mathrm{22} \mathrm{cm^{-2}}) \) & \(8.4^{+1.1}_{-1.4}\) & \(7.9^{+0.7}_{-0.8}\) & \(8.1^{+0.6}_{-0.7}\) \\
             &\( N_\mathrm{{H,3}} \) \( (10^\mathrm{22} \mathrm{cm^{-2}}) \) & \(27.5^{+3.1}_{-3.0}\) & 27.0\(\pm\)2.3 & \(27.0^{+1.9}_{-1.8}\) \\
             polconst&\(\mathrm{PD_{1}} \) (\%) & \(41.9^{+19.0}_{-15.1}\) & \(30.4^{+6.5}_{-6.1}\) & \(32.6^{+6.1}_{-5.7}\) \\
             &\(\mathrm{PA_{1}} \) ($^\circ$) & 35.6\(\pm\)9.7 & 39.3\(\pm\)5.4 & 38.3\(\pm\)4.8\\
             &\(\mathrm{PD_{2,3}} \) (\%) & 6.0\(\pm\)1.0 & 5.8\(\pm\)0.8 & 5.9\(\pm\)0.6 \\
             &\(\mathrm{PA_{2,3}} \) ($^\circ$) & -49.5\(\pm\)5.1 & -49.5\(\pm\)4.0 & -49.6\(\pm\)3.2 \\
             powerlaw&\(\Gamma\) & 2.16\(\pm\)0.12 & 1.88\(\pm\)0.08 & 1.94\(\pm\)0.07 \\
             &\(\mathrm{Norm_{1}} \)& 0.010\(\pm\)0.002 & 0.018\(\pm\)0.002 & 0.015\(\pm\)0.001 \\
             &\(\mathrm{Norm_{2+3}} \)& \(1.216^{+0.167}_{-0.114}\) & \(0.803^{+0.068}_{-0.053}\) & \(0.859^{+0.059}_{-0.048}\) \\
              gauss&\(\mathrm{E\ (keV)} \)& \(6.35^{+0.13}_{-0.11}\) & 6.39\(\pm\)0.04 & 6.39\(\pm\)0.04 \\
              &\(\mathrm{\sigma\ (keV)} \)& [0.430] & [0.430] & [0.430] \\
              &\(\mathrm{Norm} \)& 0.002 & 0.004 & 0.003 \\
        \midrule
        &\(\chi^2/\mathrm{d.o.f.} \) & 1568/1329 & 1450/1337 & 1541/1329 \\
        \bottomrule
    \end{tabular}
\caption{Spectral parameters for each observation and the combined dataset in 2--8 keV energy range obtained by the spectro-polarimetric analysis, as discussed in Section \ref{Spectral polarimetric analysis}. \(\mathrm{Norm_{2+3}}\) denotes the total \(powerlaw\) normalization of Component 2,3, as defined in Section \ref{Spectral polarimetric analysis}.}
\label{total_parameters}
\end{table*}

The \(const\) model represents cross-calibration constants. It accounts for calibration discrepancies between DUs, with DU1 fixed to unity in each observation.

We designate the first \(powerlaw\) component as Component 1 and the last two as Component 2,3. The \(tbabs\) model represents neutral hydrogen absorption, characterized by a single parameter: the equivalent hydrogen column density (\(N_{\mathrm{H}}\)). We define \(N_{\mathrm{H}}\) of the first \(tbabs\) model as \(N_{\mathrm{H,1}}\), and fix it to the typical interstellar absorption value for Vela X-1 (\(0.75\times10^{22}~\mathrm{cm}^{-2}\), see \cite{vg1981}), implying that Component 1 remains unaffected by wind clumps and is solely absorbed by the interstellar medium. The subsequent two \(tbabs\) models are fitted freely, accounting for absorption by complex clumpy winds within the Vela X-1 system. We define them as \(N_{\mathrm{H,2}}\) and \(N_{\mathrm{H,3}}\).

The \(polconst\) model, a multiplicative model describing constant polarization with energy, contains two parameters: \(\mathrm{PD}\) and \(\mathrm{PA}\). Component 1 corresponds to an independent constant polarization component, while Component 2,3 share a common constant polarization component, representing the emission under the collective effect of wind clumps. Thus, we define the following parameter sets: \(\mathrm{PD_{1}}\) and \(\mathrm{PA_{1}}\) correspond to Component 1, \(\mathrm{PD_{2,3}}\) and \(\mathrm{PA_{2,3}}\) to Component 2,3.

Additionally, we include a Gaussian component to model the $\mathrm{Fe~K}\alpha$ line at 6.4 keV, consistent with previous studies (e.g., \cite{Fuerst2013NuSTARDO}, \cite{Grinberg2017TheCA}, \cite{Diez2022ContinuumCL}, and \cite{Rahin2023ANV}). When the line parameters are left free during fitting, the best-fit solutions for both individual and combined datasets yield a line width smaller than IXPE's energy resolution in the 6--7 keV range. Since such narrow widths cannot be reliably constrained by IXPE, we fix the Gaussian sigma to 0.43 keV. This corresponds to a full width at half maximum (FWHM) of 1.02 keV, matching IXPE's FWHM resolution around 6.4 keV \citep{Weisskopf2022ImagingXP}.  

The choice of 0.43 keV for the fixed sigma is supported by previous high-resolution studies. \cite{Fuerst2013NuSTARDO} reported an orbitally averaged $\mathrm{Fe~K}\alpha$ line sigma of $\leq$0.7 keV using NuSTAR observations. \cite{Grinberg2017TheCA} found the line sigma varying between $\sim$0.5 keV and 0.8 keV at orbital phases 0.21--0.25 with Chandra data, depending on flux hardness ratios. \cite{Rahin2023ANV} observed no significant orbital dependence in the $\mathrm{Fe~K}\alpha$ line flux. Collectively, these measurements support our adoption of 0.43 keV for the sigma parameter.

Furthermore, in our fitting procedures for all datasets, we first test free fitting with unconstrained Gaussian parameters. If the fitted Norm approaches zero, we fix both sigma and Norm to zero and refit to maintain model consistency throughout this work.

For the combined dataset from both observations, with sigma fixed at 0.43 keV, the fitted line centroid lies precisely at 6.4 keV. The F-test yields a value of 38.8 with a p-value less than $1 \times 10^{-15}$, and the differences in the Akaike Information Criterion (AIC) and Bayesian Information Criterion (BIC) are $\Delta\mathrm{AIC} = -85.92$ and $\Delta\mathrm{BIC} = -75.52$, respectively. These results strongly support the inclusion of the Gaussian component. Moreover, since the Gaussian iron line is expected to be unpolarized, we also assign a \(polconst\) to the Gaussian component and fix its PD to zero \citep[see, e.g.,][]{Churazov02, Veledina24, LaMonaca24GX340}.

By freezing the \(powerlaw\) normalizations from the initial fit, convolving with the \(cglumin\) model, and refitting, we derive the unabsorbed luminosity in the 2--8 keV energy band. Subsequent analysis uses \(\mathrm{Norm_{1}}\) and \(\mathrm{Luminosity_{1}}\) for the \(powerlaw\) normalization and unabsorbed luminosity of Component 1, while \(\mathrm{Norm_{2+3}}\) and \(\mathrm{Luminosity_{2,3}}\) denote the total \(powerlaw\) normalization and unabsorbed luminosity of Component 2,3.

Using XSPEC, we perform spectral fitting on the two IXPE observations of Vela X-1 in the 2--8 keV band. Additionally, we manually merge the spectra extracted from these two observations, obtain the total exposure time using XSELECT, and subsequently fit the merged spectrum. The best-fit spectrum for the combined dataset is shown in Figure \ref{spec}. The results from fitting the datasets individually and the merged dataset are presented in Table \ref{total_parameters}.

\begin{figure}  % [t]
\centering
\includegraphics[width=0.47\textwidth]{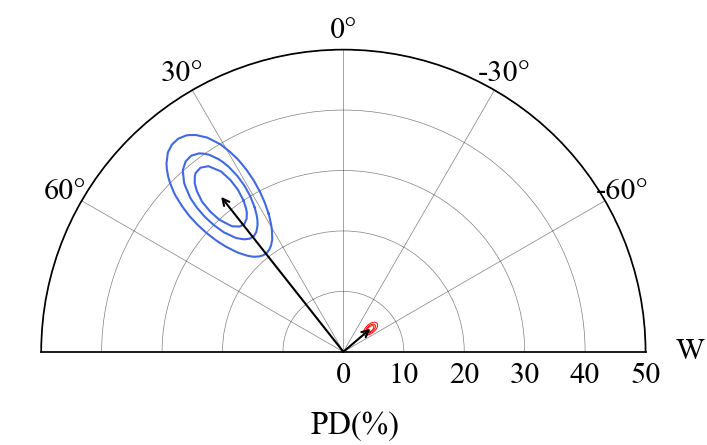}
\caption{The polarization vectors of the combined dataset of two IXPE observations. The blue contours show the polarization vectors of Component 1, while the red shows the joint polarization vectors of Component 2,3. For each set of vectors, the three contours from outward to inward correspond to the 68.30\%, 95.45\%, and 99.73\% confidence levels, respectively. Due to the narrow uncertainty ranges, the three red contours are visually indistinguishable at this scale.}
\label{contour}
\end{figure}

As shown in Table \ref{total_parameters}, we find that both the second observation and the combined dataset exhibit significantly high \(\mathrm{PD_{1}}\) values, with significance levels of \(\sim5\sigma\). Additionally, \(\mathrm{PA_{1}}\) and \(\mathrm{PA_{2,3}}\) exhibit a \(90^\circ\) offset. The confidence contours for the polarization measurements were created by the \textit{steppar} command in XSPEC and are shown in Figure \ref{contour} for the combined dataset.

We also attempt to separately compute the polarization of Component 1, Component 2, and Component 3 in the 2--8 keV energy band using XSPEC. We obtained \(\mathrm{PD_{1}} = 14.6 \pm 9.0\%\), \(\mathrm{PA_{1}} = 36.5 \pm 17.7^\circ\), \(\mathrm{PD_{2}} = 1.4 \pm 3.0\%\) (consistent with zero), with \(\mathrm{PA_{2}}\) unconstrained, \(\mathrm{PD_{3}} = 10.9 \pm 2.1\%\), and \(\mathrm{PA_{3}} = -49.9 \pm 5.6^\circ\). The flux of Component 2 dominates in the 2.5--4 keV range, as shown in Figure \ref{spec}. Moreover, the PA values of Component 1 and Component 3 exhibit an orthogonal trend, and their flux becomes comparable in the 3--4 keV energy band. Therefore, these can explain the zero degree of polarization around 3.5 keV reported in \cite{Forsblom2023IXPEOO}. The 90\(^\circ\) swing in PA between 2--3 keV and 4--8 keV mentioned in \cite{Forsblom2023IXPEOO} can also be explained by the polarization behavior of Component 1 and Component 3.

\setlength{\tabcolsep}{3pt}
\begin{table*}[t]
\centering
    \begin{tabular}{c|ccc|ccccc|cc}
        \toprule
            \textbf{MJD} &   \(\bm{\mathrm{Luminosity_{1}}}\) & \(\bm{\mathrm{PD_{1}}}\) & \(\bm{\mathrm{PA_{1}}}\)  & \( \bm{N_\mathrm{{H,2}}} \)&\( \bm{N_\mathrm{{H,3}}} \)&\(\bm{\mathrm{Luminosity_{2,3}}}\) & \(\bm{\mathrm{PD_{2,3}}}\) & \(\bm{\mathrm{PA_{2,3}}}\) & \(\bm{\Gamma}\) &\( \bm{\chi^2/} \)\\
       &   \( \bm{(10^\mathrm{35} \mathrm{erg\ s^{-1}})} \) & \textbf{(\%)} & \textbf{($^\circ$)} &\( \bm{(10^\mathrm{22} \mathrm{cm^{-2}})} \)&\( \bm{(10^\mathrm{22} \mathrm{cm^{-2}})} \)&\( \bm{(10^\mathrm{35} \mathrm{erg\ s^{-1}})} \) & \textbf{(\%)} & \textbf{($^\circ$)} & & \(\bm{\mathrm{d.o.f.}}\)\\
        \midrule
        (a) & 0.05& 32.1\(\pm\)18.1 & 51.7\(\pm\)16.1 & \(13.6^{+1.4}_{-1.3}\) & \(34.8^{+5.5}_{-3.6}\) & \(2.48\) & 6.0\(\pm\)1.3 & -44.0\(\pm\)6.1 &2.23& 1483/1337 \\
        (b) &   0.17& 40.7\(\pm\)16.8 & 24.1\(\pm\)11.8 & 8.8\(\pm\)0.6 & \(37.9^{+4.6}_{-3.9}\) & \(4.11\) & 5.9\(\pm\)1.7 & -59.2\(\pm\)8.4 &2.28& 1383/1337 \\
        (c) &  0.53& 36.5\(\pm\)11.7 & 41.9\(\pm\)9.4 & \(5.4^{+0.9}_{-1.1}\) & \(26.5^{+4.6}_{-4.4}\) & \(4.82\) & 7.1\(\pm\)2.0 & -44.8\(\pm\)8.1 &1.95& 1347/1337 \\
        (d) &   0.20 & 33.2\(\pm\)8.4 & 44.8\(\pm\)7.3 & \(8.0^{+1.7}_{-1.5}\) & \(29.0^{+5.1}_{-3.0}\) & \(2.28\) & 7.2\(\pm\)1.7 & -53.1\(\pm\)6.7 &1.68& 1365/1337 \\
        (e) &  0.08& 45.4\(\pm\)17.1 & 20.4\(\pm\)10.8 & 12.9\(\pm\)0.8 & \(35.5^{+5.1}_{-3.9}\) & \(3.76\) & 4.1\(\pm\)1.0 & -52.3\(\pm\)7.3 &2.19& 1397/1337 \\
        \bottomrule
    \end{tabular}
    \caption{Spectral parameters of different time intervals in 2--8 keV energy band obtained by the spectro-polarimetric analysis. In all time intervals, \( N_\mathrm{H,1} \) is fixed to the interstellar absorption value for Vela X-1 (\(0.75\times10^{22}~\mathrm{cm}^{-2}\)). The unabsorbed luminosity of Component 1 and the total unabsorbed luminosity of Component 2,3 are also shown. (a)--(e) respectively correspond to the MJD time intervals: (a) 59684.8--59687, (b) 59689.5--59690.5, (c) 59914--59915, (d) 59915--59917.2, and (e) 59917.2--59919.4.  }
\label{d_parameters}
\end{table*}

\begin{figure*}[t]  % [t]
\centering
\vspace{0pt}
%\hspace{0.1\textwidth}
\includegraphics[width=0.3\textwidth]{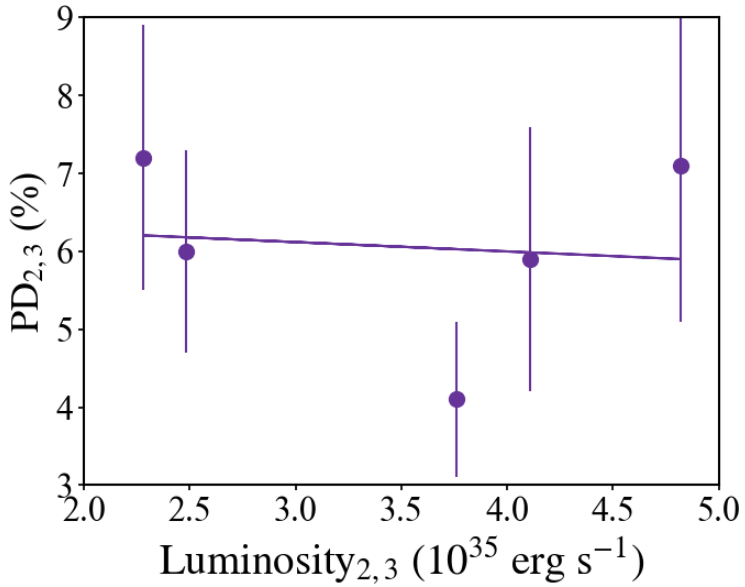}%
\hspace{20pt}
\includegraphics[width=0.3\textwidth]{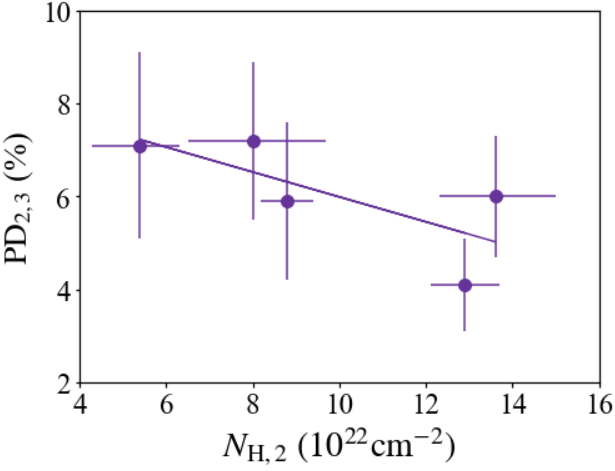}%
\hspace{20pt}
\includegraphics[width=0.3\textwidth]{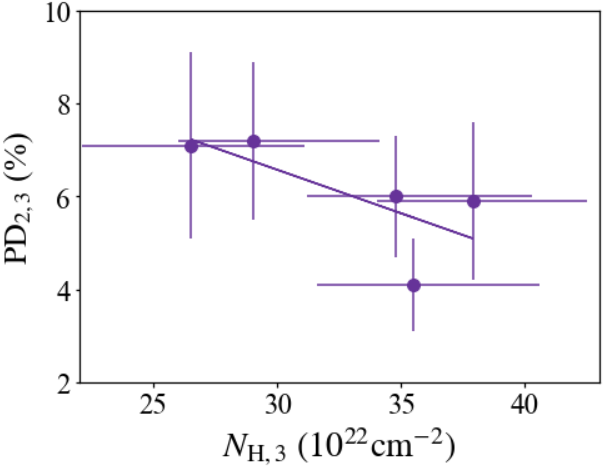}
\caption{Correlation of parameters in Table \ref{d_parameters}. The left, middle and right panel respectively shows \(\mathrm{PD_{2,3}} \) versus \(\mathrm{Luminosity_{2,3}}\), \(N_\mathrm{{H,{2}}}\) and \(N_\mathrm{{H,{3}}}\).}
\label{CA2}
\end{figure*}

\subsection{Time-resolved polarimetric analysis}
\label{Time-resolved and Correlation analysis}

To investigate temporal variations in spectral parameters and polarization properties, we partition the two observational datasets into five epochs: (a) to (e), as shown in Figure \ref{lc}. Epochs (b) and (c) correspond to bright states within the orbits. To ensure sufficient photon counts, we set the duration of (b) and (c) to 1 day each and allocate 2.2 days to each of the remaining epochs. The final temporal segmentation is: (a) 59684.8--59687, (b) 59689.5--59690.5, (c) 59914--59915, (d) 59915--59917.2 and (e) 59917.2--59919.4. We subsequently fit the data from these five epochs using a methodology consistent with our previous analysis. The corresponding \( N_\mathrm{H} \), unabsorbed luminosity, polarization parameters, and \(\Gamma\) in the 2--8 keV energy range are tabulated in Table \ref{d_parameters}.

We observe that \( N_\mathrm{H,2} \) and \( N_\mathrm{H,3} \) fluctuate around \(\sim10\times10^{22}\ \mathrm{cm}^{-2}\) and \(\sim30\times10^{22}\ \mathrm{cm}^{-2}\), respectively, indicating absorption by a combination of distinct wind clumps. Additionally, across all epochs, \(\mathrm{PA_{1}}\) and \(\mathrm{PA_{2,3}}\) remain nearly orthogonal within errors or exhibit signs of orthogonality.

\(\mathrm{PD_{1}}\) consistently shows higher values in all epochs. Conversely, \(\mathrm{PD_{2,3}}\) is generally lower, consistent with the polarization values of Vela X-1 derived in previous research \citep{Forsblom2023IXPEOO, Forsblom2025RevealingTO} and typical IXPE observational values for XRPs (for a review, see \cite{Poutanen2024XrayPO}).

During epochs (c) and (d), we measure \(\mathrm{PD_{1}}\) values of \(36.5 \pm 11.7\%\) and \(33.2 \pm 8.4\%\) with significances of \(>3\sigma\). Additionally, \(\mathrm{PD_{2,3}}\) reaches its highest central values in epochs (c) and (d) at \(7.1\%\) and \(7.2\%\), corresponding to significances of \(\sim3.6\sigma\) and \(\sim4.2\sigma\) respectively. Simultaneously, we find that \( N_\mathrm{H,2} \), \( N_\mathrm{H,3} \), and the photon index reach their lowest values during (c) and (d) across all epochs.

To further investigate factors influencing Vela X-1's \(\mathrm{PD}\), we performed correlation analyses. Due to insufficient statistics \(\mathrm{PD_{1}}\) in epochs (a) and (b), we merely analyze the correlations between \(\mathrm{PD_{2,3}}\) and other spectral parameters. We find no correlation between \(\mathrm{PD_{2,3}}\) and \(\mathrm{Luminosity_{2,3}}\), but observe potential negative correlations with \( N_\mathrm{H,2} \) and \( N_\mathrm{H,3} \), showing Pearson coefficients of \(-0.740\) and \(-0.716\) (\(p\)-values = 0.153 and 0.174, respectively). These correlation results are shown in Figure \ref{CA2}.

\subsection{Phase-resolved analysis of different components}
\label{Phase-resolved polarimetric analysis}

\begin{figure*}[t]  % [t] 
\centering
\vspace{0pt}
\hspace{0pt}
\includegraphics[width=0.368\textwidth]{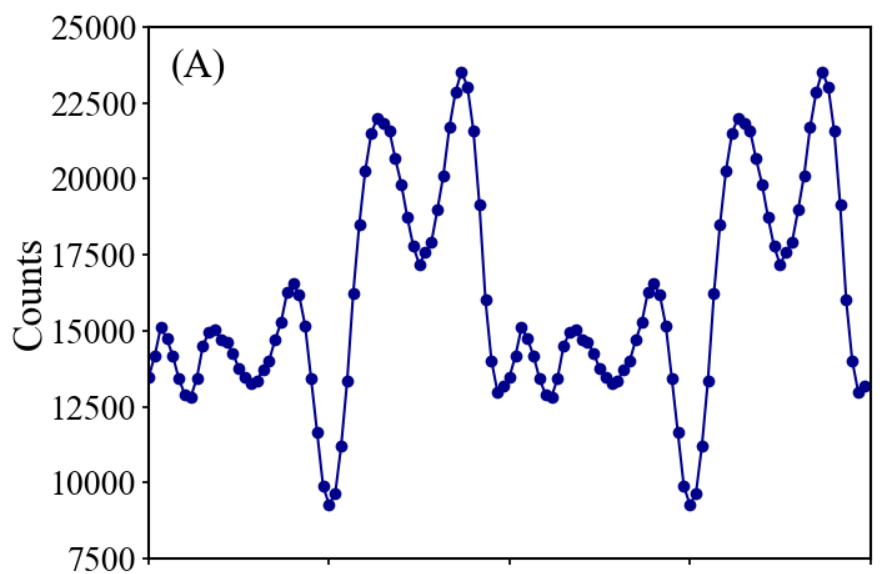}%
\hspace{2pt}
\includegraphics[width=0.35\textwidth]{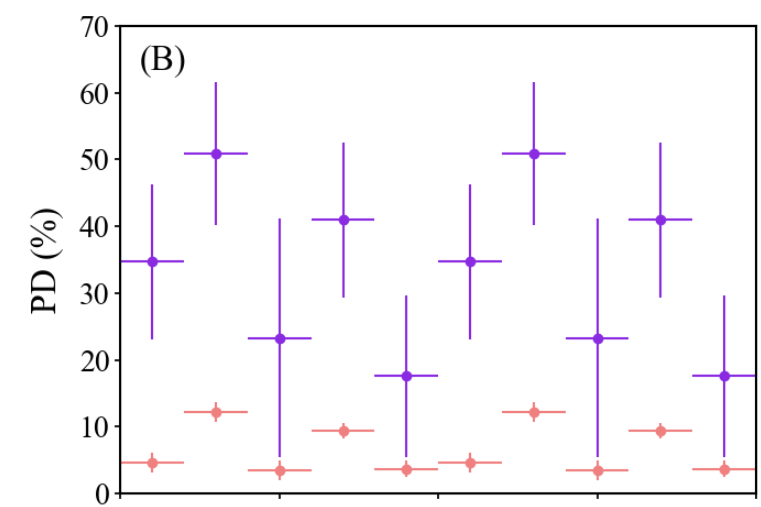}\\
\hspace{13pt}
\includegraphics[width=0.357\textwidth]{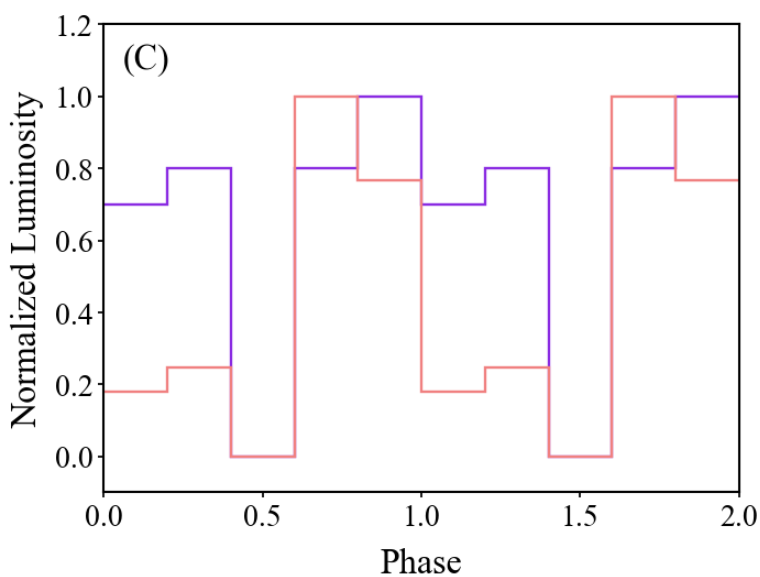}%
\hspace{-3pt}
\includegraphics[width=0.365\textwidth]{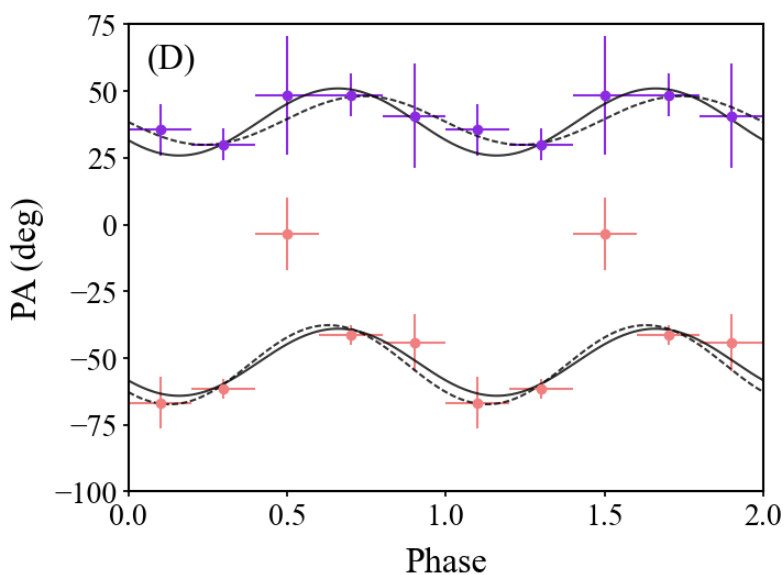}%
\caption{Results of the phase-resolved spectro-polarimetric analysis, using the combined dataset of two IXPE observations in 2--8 keV energy band. (A): The overall pulse profiles. (B): The phase-resolved \(\mathrm{PD_{1}} \) and \(\mathrm{PD_{2,3}} \). The purple line shows Component 1, and the light coral line shows Component 2,3. (C): Same colors indicate the variation of \(\mathrm{Luminosity_{1}}\) and \(\mathrm{Luminosity_{2,3}}\). (D): Same colors indicate the phase-resolved \(\mathrm{PA_{1}} \) and \(\mathrm{PA_{2,3}} \). The black solid curve shows the best-fit RVM for simultaneous fitting of $\mathrm{PA_1}$ and $\mathrm{PA_{2,3}}$. The black dashed curve represents the best-fit RVM for separate fitting of $\mathrm{PA_1}$ and $\mathrm{PA_{2,3}}$.}
\label{Phase PD}
\end{figure*}

The phase offset between the pulse profiles of the first and second observations was determined from cross-correlation (using the implementation provided by the Python library NumPy). Then, we divide the combined dataset of two observations into multiple phase intervals (each of width 0.2) and fit them with the method discussed in Section \ref{Spectral polarimetric analysis} in the 2--8 keV energy band using XSPEC. The derived parameters are shown in Table \ref{phase_parameters} and Figure \ref{Phase PD}. 

We find that in the phase interval 0.2--0.4, \(\mathrm{PD_{1}}\) is detected at \(50.9\%\) with a significance of \(\sim5\sigma\). Additionally, the phase intervals 0.0--0.2 and 0.6--0.8 also show high \(\mathrm{PD_{1}}\) values of \(34.7 \pm 11.6\%\) and \(41.0 \pm 11.6\%\), corresponding to significances of \(>3\sigma\) and \(\sim4\sigma\) respectively.

We also find that \(\mathrm{Luminosity_{1}}\) and \(\mathrm{Luminosity_{2,3}}\) pulsate synchronously and vary in sync with the overall pulse profile, as shown in Figure \ref{Phase PD}(A) and (B). Additionally, Figure \ref{Phase PD}(B) reveals that Component 1 and Component 2,3 contribute to different parts of the total pulse profile, indicating that their emission regions and patterns may differ.

Additionally, Figure \ref{Phase PD}(D) shows that both \(\mathrm{PA}_1\) and \(\mathrm{PA}_{2,3}\) exhibit the same variation trend with phase and follow a sinusoidal pattern. The only exception is \(\mathrm{PA}_{2,3}\) in the 0.4--0.6 phase interval, which may result from the difficulty in constraining the PA due to the low significance of \(\mathrm{PD}_{2,3}\), or may indicate more complex polarization properties within the system. Overall, we can fit the Rotating Vector Model (RVM) to the pulse-phase dependent PA using the Rotating Vector Model (RVM) with the affine-invariant Markov Chain Monte Carlo (MCMC) ensemble sampler \textit{emcee} in Python \citep{ForemanMackey2012emceeTM}.

Previously, the RVM has been successfully applied to the phase-dependent PA variation in several X-ray accreting pulsars \citep{{Doroshenko2022DeterminationOX},{Tsygankov2022TheXP}, {Mushtukov2023XrayPO}, {Tsygankov2023XrayPG}, {Malacaria2023APO}, {Doroshenko2023ComplexVI}, {Suleimanov2023XrayPO}, {Heyl2024ComplexRD}, {Zhao2024PolarizationPO}, {Poutanen2024StudyingGO}, {Forsblom2024ProbingTP}, {Forsblom2025RevealingTO}}. In the RVM, if radiation escapes in the O-mode, PA can be described by the following equation (see Eq. (30) in \cite{Poutanen2020RelativisticRV}):

\begin{equation} \label{eq:pa_rvm}
\tan (\mbox{PA}\!-\!\chi_{\rm p})\!=\! \frac{-\sin \theta\ \sin (\phi-\phi_0)}
{\sin i_{\rm p} \cos \theta\!  - \! \cos i_{\rm p} \sin \theta  \cos (\phi\!-\!\phi_0) } ,
\end{equation}

\setlength{\tabcolsep}{2pt}
\begin{table*}[t]
\centering
    \begin{tabular}{c|ccc|ccccc|cc}
        \toprule
            \textbf{Phase} &  \(\bm{\mathrm{Luminosity_{1}}}\) & \(\bm{\mathrm{PD_{1}}}\) & \(\bm{\mathrm{PA_{1}}}\)  & \( \bm{N_\mathrm{{H,2}}} \)&\( \bm{N_\mathrm{{H,3}}} \)&\(\bm{\mathrm{Luminosity_{2,3}}}\) & \(\bm{\mathrm{PD_{2,3}}}\) & \(\bm{\mathrm{PA_{2,3}}}\) & \(\bm{\Gamma}\) &\( \bm{\chi^2/} \)\\
        &  \( \bm{(10^\mathrm{35} \mathrm{erg\ s^{-1}})} \) & \textbf{(\%)} & \textbf{($^\circ$)} &\( \bm{(10^\mathrm{22} \mathrm{cm^{-2}})} \)&\( \bm{(10^\mathrm{22} \mathrm{cm^{-2}})} \)&\( \bm{(10^\mathrm{35} \mathrm{erg\ s^{-1}})} \) & \textbf{(\%)} & \textbf{($^\circ$)} & & \(\bm{\mathrm{d.o.f.}}\)\\
        \midrule
        0.0--0.2 & 0.17 & 34.7\(\pm\)11.6  & 35.5\(\pm\)9.7 & \(8.1^{+1.7}_{-1.5}\) & \(37.9^{+4.6}_{-3.9}\) & 2.84 & 4.6\(\pm\)1.5 & -66.8\(\pm\)9.7 & 1.91 & 1460/1337 \\
        0.2--0.4 & 0.18 & 50.9\(\pm\)10.7 & 29.9\(\pm\)6.0 & \(8.7^{+1.5}_{-1.8}\) & \(28.9^{+5.5}_{-4.6}\) & 2.94 & 12.2\(\pm\)1.5 & -61.6\(\pm\)3.6 & 1.88 & 1311/1337 \\
        0.4--0.6 & 0.10 & 23.3\(\pm\)17.9 & 48.5\(\pm\)22.3 & \(6.9^{+2.0}_{-1.6}\) & \(24.0^{+5.1}_{-2.5}\) & 2.57 & 3.4\(\pm\)1.5 & -3.6\(\pm\)13.5 & 2.17 & 1310/1337 \\
        0.6--0.8 & 0.18 & 41.0\(\pm\)11.6 & 48.5\(\pm\)8.1 & 8.9\(\pm\)0.9 & \(30.3^{+3.8}_{-2.9}\) & 4.07 & 9.4\(\pm\)1.2 & -41.4\(\pm\)3.7 & 2.13 & 1475/1337 \\
        0.8--1.0 & 0.20 & 17.6\(\pm\)12.1 & 40.7\(\pm\)19.6 & \(7.7^{+1.5}_{-1.4}\) & \(25.6^{+5.9}_{-3.4}\) & 3.72 & 3.7\(\pm\)1.3 & -44.3\(\pm\)10.7 & 1.96 & 1438/1337 \\
        \bottomrule
    \end{tabular}
    \caption{Pulse phase dependence of spectral parameters in 2--8 keV obtained by the spectro-polarimetric analysis, using the combined dataset of two IXPE observations. In all phase intervals, \( N_\mathrm{H,1} \) is fixed to the interstellar absorption value for Vela X-1 (\(0.75\times10^{22}~\mathrm{cm}^{-2}\)).}
\label{phase_parameters}
\end{table*}

%%%%%%%%%%%%%%%%%%%%%%%%%%%%
%%%%%%%%%%%%%%%%%%%%%%%%%%%%

\noindent where $i_p$ is the inclination angle between the pulsar spin axis and the line of sight, $\theta$ is the magnetic obliquity (angle between the magnetic dipole and spin axis), $\chi_p$ is the pulsar position angle (measured from North to East), and $\phi_0$ is the phase when the magnetic north pole crosses the observer's meridian. When radiation escapes predominantly in the X-mode, the pulsar position angle becomes $\chi_p \pm 90^\circ$.

We first fit the RVM to $\mathrm{PA_1}$ and $\mathrm{PA_{2,3}}$ separately and constrain $i_p$ to 73$^\circ-107^\circ$ based on Vela X-1's orbital inclination limit (for a review, see \cite{Kretschmar2021RevisitingTA}). We show covariance plots in Figures \ref{mcmch} and \ref{mcmcl} and list resulting parameters in the first and second row of Table \ref{RVM}. Both solutions agree within errors. 

Subsequently, we simultaneously fit the RVM to $\mathrm{PA_1}$
and $\mathrm{PA_{2,3}}$. Prior to fitting, we applied an additional 90$^\circ$ rotation to the PA value predicted by the RVM for $\mathrm{PA_1}$. Figure \ref{mcmc1} shows covariance plots, while the third row of Table \ref{RVM} presents parameters. The individual and joint fitting results are mutually consistent and align with the fitting outcomes obtained by \citet{Forsblom2025RevealingTO} using only high-energy segment data, showing a small magnetic obliquity angle $\theta$ and weak RVM constraints on $i_p$. We attempt to fix $i_p$ to our obtained best-fit value and perform the fit again, but this can not yield any improvement in the fit.

\begin{table*}[t]
\centering
    \begin{tabular}{ccccc}
        \toprule
             &   \(\bm{i_p}\) & \(\bm{\theta}\)  & \( \bm{\chi_p} \)&\( \bm{\phi_0/2\pi} \)\\
        &   \textbf{(deg)} & \textbf{(deg)} &\textbf{(deg)} &\\
        \midrule
        Component 1 & \(88.26^{+10.08}_{-10.04}\) & \(9.12^{+5.69}_{-5.36}\) & \(128.98^{+4.26}_{-4.29}\) & \(0.99^{+0.14}_{-0.14}\) \\
        Component 2,3& \(90.18^{+8.92}_{-11.28}\) & \(14.81^{+4.61}_{-3.95}\) & \(127.49^{+2.67}_{-2.70}\) & \(0.88^{+0.06}_{-0.04}\) \\
        Total & \(89.30^{+9.45}_{-10.69}\) & \(12.56^{+3.19}_{-2.82}\) & \(128.41^{+2.16}_{-2.22}\) & \(0.91^{+0.06}_{-0.05}\) \\
        \bottomrule
    \end{tabular}
    \caption{Best-fit RVM parameters for separate fitting and simultaneous fitting of $\mathrm{PA_1}$ (corresponding to Component 1) and $\mathrm{PA_{2,3}}$ (corresponding to Component 2,3).}
\label{RVM}
\end{table*}

%%%%%%%%%%%%%%%%%%%%%%%%%%%%%%%%%%%%%%%%%%%%%%%%%%%%%%%%%%%%%%%%%%%%%%%%%%%
\section{DISCUSSION} \label{sec:discussion}

In previous studies of Vela X-1 using a similar three power-law component model \citep{MartinezNunez2014TheAE} to fit the \textit{XMM-Newton} data in the 0.5--10 keV energy range, the authors proposed that the two most variable components across epochs correspond to two aspects of the emission from the NS surface. These exhibited characteristic \(N_\mathrm{{H}}\) of $\sim 30\times10^{22}\ \mathrm{cm^{-2}}$ and $\sim 4\times10^{22}\ \mathrm{cm^{-2}}$. Another power-law might correspond to a soft excess below 3 keV caused by emisedsion from photoionized or collisionally heated diffuse gas, or thermal emission from the neutron star surface \citep{Hickox2024}, with a \(N_\mathrm{{H}}\) comparable to the interstellar absorption value, interpreted as a light echo in the distant stellar wind.

In our study, the 2--8 keV spectrum of Vela X-1 can also be fitted by the three power-law model. Figure \ref{Phase PD}(C) shows that the unabsorbed luminosities of Component 1 and Component 2,3 both vary approximately synchronously with the pulse. More importantly, the PA of these three components pulsates synchronously with the pulse phase, and the RVM fitting parameters are consistent with previous studies of Vela X-1 \citep{Forsblom2025RevealingTO}. This evidence implies that Component 1, like Component 2,3, originates from the polar cap emission of the neutron star rather than from the distant stellar wind.

Figure \ref{Phase PD}(B) reveals that Component 2,3 primarily contribute to the first minor peak of the total pulse profile, while Component 1 exhibits a relatively uniform contribution across the entire pulse profile. This implies distinct emission regions between the components. Given that Component 1 is only subject to interstellar absorption, we speculate that it represents emission from the surface layer of the accretion mound. Photons escaping this region do not traverse the main body of the accretion mound nor the accretion flow, but instead pass only through the uniform, optically thin stellar wind surrounding the neutron star periphery, experiencing minor absorption in distant media.

For neutron stars with strong magnetic fields in XRPs, the medium undergoes vacuum birefringence in the intense magnetic field, forming two polarization states: the O-mode and the X-mode \citep{Gnedin1974}. Below the cyclotron resonance scattering energy (a fundamental energy around 25 keV and a first harmonic energy near 50 keV for Vela X-1, \cite{Staubert2018CyclotronLI}), the opacity of the X-mode is much lower than that of the O-mode \citep{Lai2003a}. Therefore, the expected intrinsic radiation is predicted to be predominantly polarized in the X-mode, with predicted PD exceeding 30\% across various models. Under high mass accretion rates, the PD may approach 80\%. (for a review, see \cite{Poutanen2024XrayPO}).

Subsequently, this strongly polarized intrinsic radiation traverses the uniform, optically thin stellar wind surrounding the neutron star magnetosphere. For a uniform, spherically symmetric wind, the optical depth is estimated as \(\tau_{\mathrm{T}} \simeq 2 \times 10^{-4} \dot{M}_8 \, a_{12}^{-1} \, v_{x8}^{-1}\) \citep{Kallman2015XRAYPF}, where \(\dot{M}_8\) is the mass-loss rate in units of \(10^{-8} M_{\odot} \, \mathrm{yr}^{-1}\), \(a_{12}\) is the separation distance in \(10^{12} \, \mathrm{cm}\), and \(v_{x8}\) is the wind velocity at the X-ray source in \(10^3 \, \mathrm{km \, s^{-1}}\). For Vela X-1, adopting typical parameters \(\dot{M}_8 = 1 \times 10^{-6}\) \citep{GmenezGarca2016MeasuringTS}, \(a_{12} = 3.55\) \citep{Quaintrell2003TheMO}, and an upper limit on orbital wind velocity \(v_{x8} = 0.6\) \citep{Kretschmar2021RevisitingTA}, we derive \(\tau_{\mathrm{T}} \sim 0.07\). At such a low optical depth, the PD remains nearly unchanged after scattering \citep{Suleimanov2023XrayPO}. 

This naturally explains the high PD of Component 1 we obtain. As shown in Table \ref{phase_parameters}, \(\mathrm{PD_{1}}\) is detected with high significance (\(\sim5\sigma\)) at approximately 50\% (\(50.9 \pm 10.7\%\)). This marks the first detection of such highly polarized neutron star emission in an XRP with high significance.

For Component 2,3, considering their absorption by higher column densities, synchronous pulsation with Component 1, and primary contribution to the first minor peak of the total pulse profile, they likely represent neutron star emission originating from the high side of the accretion mound accumulation along the accretion flow direction. Component 2,3 may be absorbed by clumps of varying thicknesses, resulting in differences in column density. This implies structural differences in the accretion geometry on the neutron star surface, which further signifies variations in accretion rate.

The slightly absorbed Component 2 is only partially deflected and mixes with the original radiation, resulting in PD$_2$ = 0. In contrast, Component 3, which passes through a denser region, is almost completely deflected, with PA$_3$ showing an orthogonal trend relative to PA$_1$. This can explain the 90$^\circ$ swing in PA above and below 3.5 keV mentioned in \cite{Forsblom2023IXPEOO}. The potential correlation between PD and column density suggests that vacuum resonance may occur within the accretion mound \citep{Gnedin1978TheEO}. In this resonance, the contributions of the plasma and magnetized vacuum to the dielectric tensor cancel each other out, leading to rapid conversion between the X-mode and O-mode. If the region where final scattering occurs is also located in this area, significant Faraday depolarization may arise \citep{Doroshenko2022DeterminationOX}. However, this requires the accretion mound to be sufficiently dense and accumulated in a region with a strong magnetic field \citep{Pavlov1979InfluenceOV}. In any case, determining the specific physical mechanism will require future higher-precision polarization observations of Vela X-1.

The enhanced X-ray Timing and Polarimetry mission (eXTP) scheduled for launch in early 2030 offers the capability to perform high-resolution spectral, timing, and polarimetric observations within the 0.5–-10 keV energy range \citep{Ge2025PhysicsOS}. Therefore, we anticipate that eXTP will further validate the high polarization characteristics of Vela X-1 in this energy band and help us reach more definitive conclusions.

\section{SUMMARY}
\label{sec:summary}

The results of our study can be summarized as follows:

1.The IXPE spectrum of Vela X-1 is modeled by three power-law components with distinct equivalent hydrogen column densities and a Gaussian line at $\sim$6.4 keV. Component 1 exhibits a constant polarization state with solely interstellar absorption. Component 2,3 share a constant polarization state but experience complex stellar wind absorption. These polarization states are mutually orthogonal.

2.The polarization state corresponding to the first power-law component exhibits a high PD (\(\sim\)30–50\%) in time-averaged, time-resolved, and phase-resolved analyses, exceeding values commonly observed in HMXBs. Phase-resolved analysis measures $\mathrm{PD_{1}} \sim 50\%$ at $\sim5\sigma$ significance. In contrast, \(\mathrm{PD_{2,3}}\) shows a lower PD (\(\sim\)5\%).
    
3.Time-resolved analysis reveals that across all epochs, \(\mathrm{PA_{1}}\) and \(\mathrm{PA_{2,3}}\) remain nearly orthogonal within errors or exhibit signs of orthogonality. Additionally, when the photon index reaches its minimum, the detection significance of \(\mathrm{PD_{1}}\) increases, while \(\mathrm{PD_{2,3}}\) attains its maximum central values. 

4.Correlation analysis reveals potential negative correlations between $\mathrm{PD_{2,3}}$ and column density. No correlation exists between PD and Luminosity.

5.We apply the RVM to fit \(\mathrm{PA_{1}}\) and \(\mathrm{PA_{2,3}}\). By imposing a 90$^\circ$ offset on the pulsar position angle of Component 1, the derived pulsar geometric parameters were consistent with those obtained from fitting \(\mathrm{PA_{2,3}}\).

6.Due to its interstellar-level absorption column density and pulsating behavior, we conclude that Component 1 originates from the radiation emitted from the accretion mound surface that is not obscured by the stellar wind. This radiation is dominated by a single polarization state in the strong magnetic field environment, with a high PD. Its polarization remains largely unchanged after traversing the optically thin outer stellar wind. This marks the first detection of such highly polarized neutron star emission in an XRP with high significance.

7.The low PD of Component 2,3 likely stems from complex physical processes in the vicinity of the accretion mound. Determining the specific mechanism will require future higher-precision polarization observations of Vela X-1.

\section*{Acknowledgments}
This work is supported by National Key R\&D Program of China (grant No. 2023YFE0117200), and National Natural Science Foundation of China (grant No. 12373041, No. 12422306 and No. 12373051), and Bagui Scholars Program (XF). This work is also supported by the Guangxi Talent Program (“Highland of Innovation Talents”).
This research used data products provided by the IXPE Team (MSFC, SSDC, INAF, and INFN) and distributed with additional software tools by the High-Energy Astrophysics Science Archive Research Center (HEASARC), at NASA Goddard Space Flight Center (GSFC). The Imaging X-ray Polarimetry Explorer (IXPE) is a joint US and Italian mission.  
FLM contribution is supported by the Italian Space Agency (Agenzia Spaziale Italiana, ASI) through contract ASI-INAF-2022-19-HH.0, by the Istituto Nazionale di Astrofisica (INAF) in Italy, and partially supported by MAECI with grant CN24GR08 “GRBAXP: Guangxi-Rome Bilateral Agreement for X-ray Polarimetry in Astrophysics”.

\clearpage
\appendix

\setcounter{figure}{0}
\centering
\renewcommand{\thefigure}{A\arabic{figure}}
\begin{figure*}[hbt]
\centering
\includegraphics[width=0.6\textwidth]{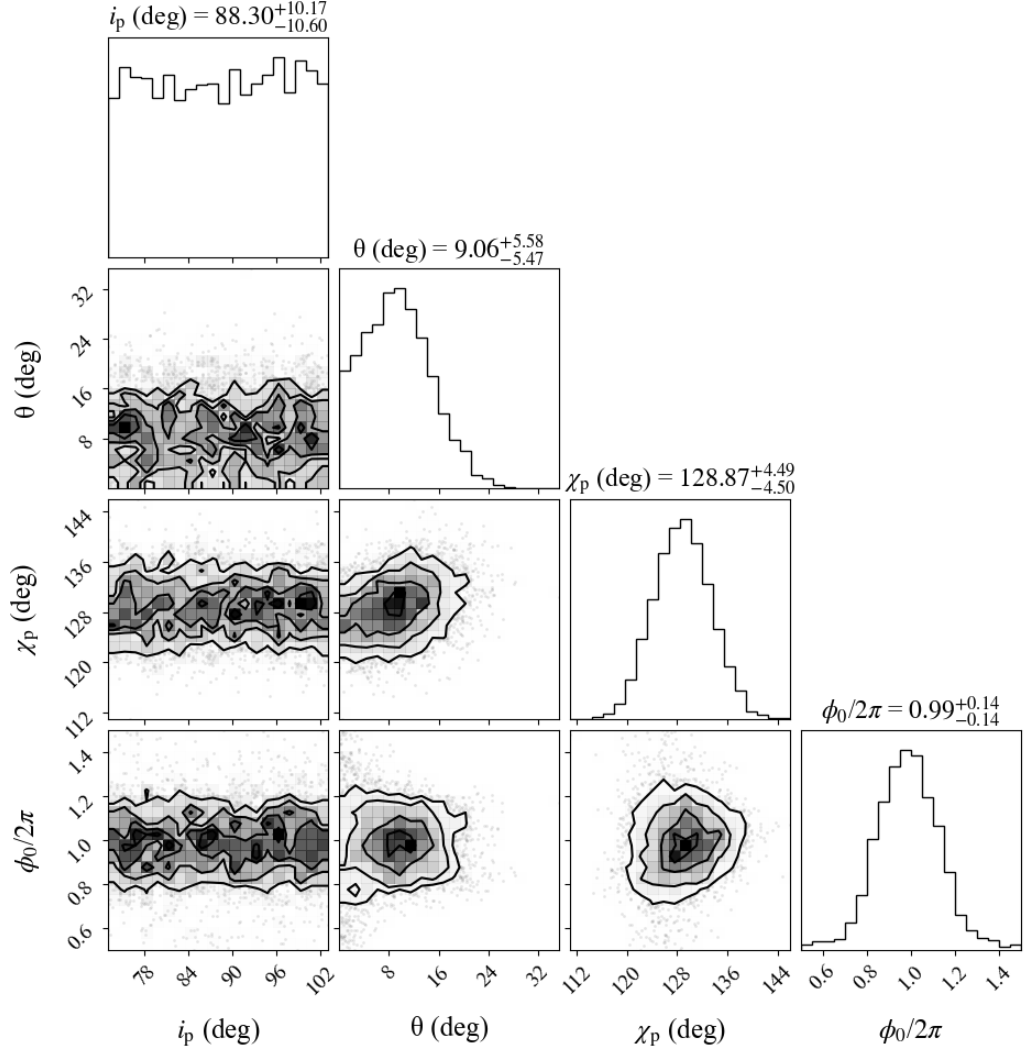}
\caption{Corner plot of the posterior distribution for the parameters of RVM fitted to phase-resolved \(\mathrm{PA_{1}}\), as discussed in Section \ref{Phase-resolved polarimetric analysis}. The range of \(i_p\) is confined to 73$^\circ$–107$^\circ$. The two-dimensional contours correspond to 68.3\%, 95.45\% and 99.73\% confidence levels. The histograms show the normalized one-dimensional distributions for a given parameter derived from the posterior samples.}
\label{mcmch}
\end{figure*}

\begin{figure*}[hbt]
\centering
\includegraphics[width=0.6\textwidth]{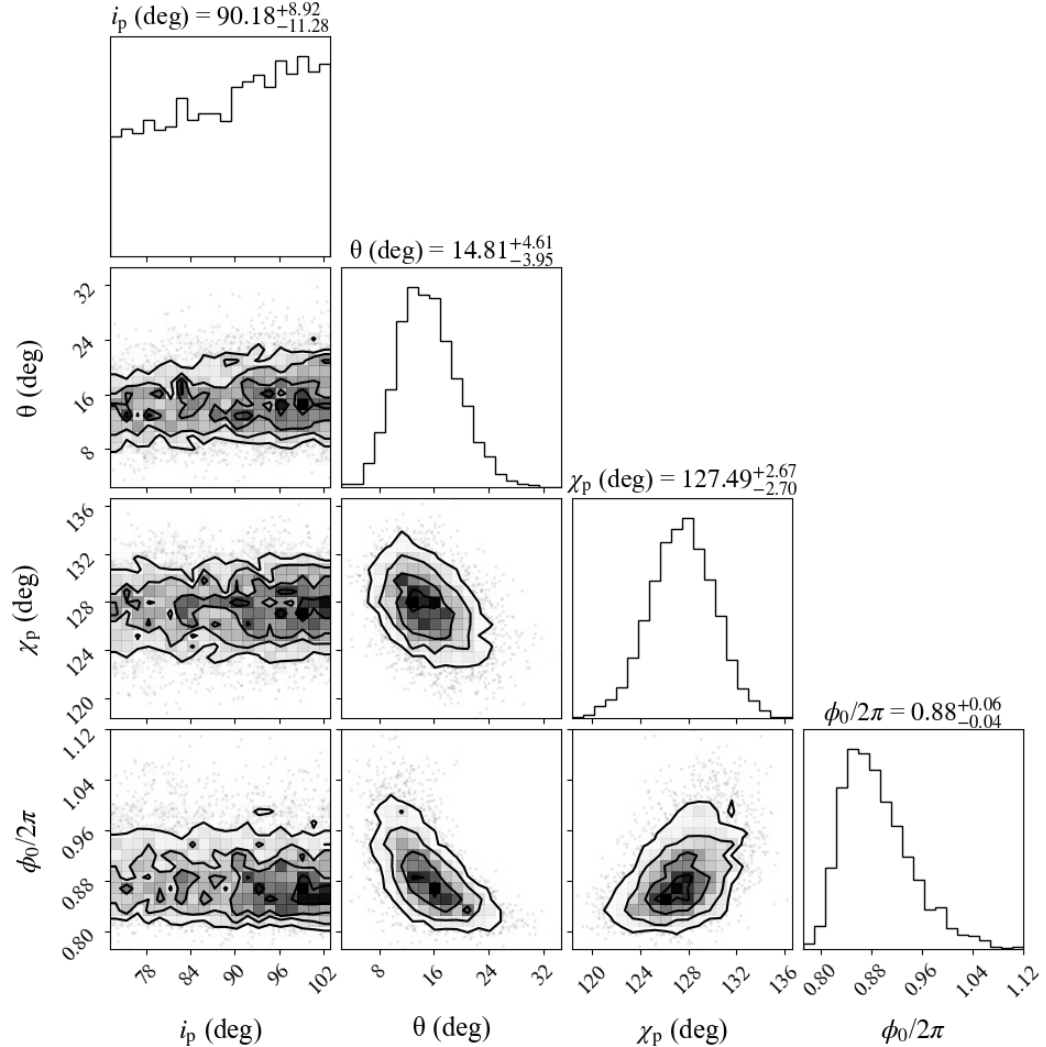}
\caption{Same as Figure \ref{mcmch}, but for phase-resolved \(\mathrm{PA_{2,3}}\)}
\label{mcmcl}
\end{figure*}

\begin{figure*}[hbt]
\centering
\includegraphics[width=0.6\textwidth]{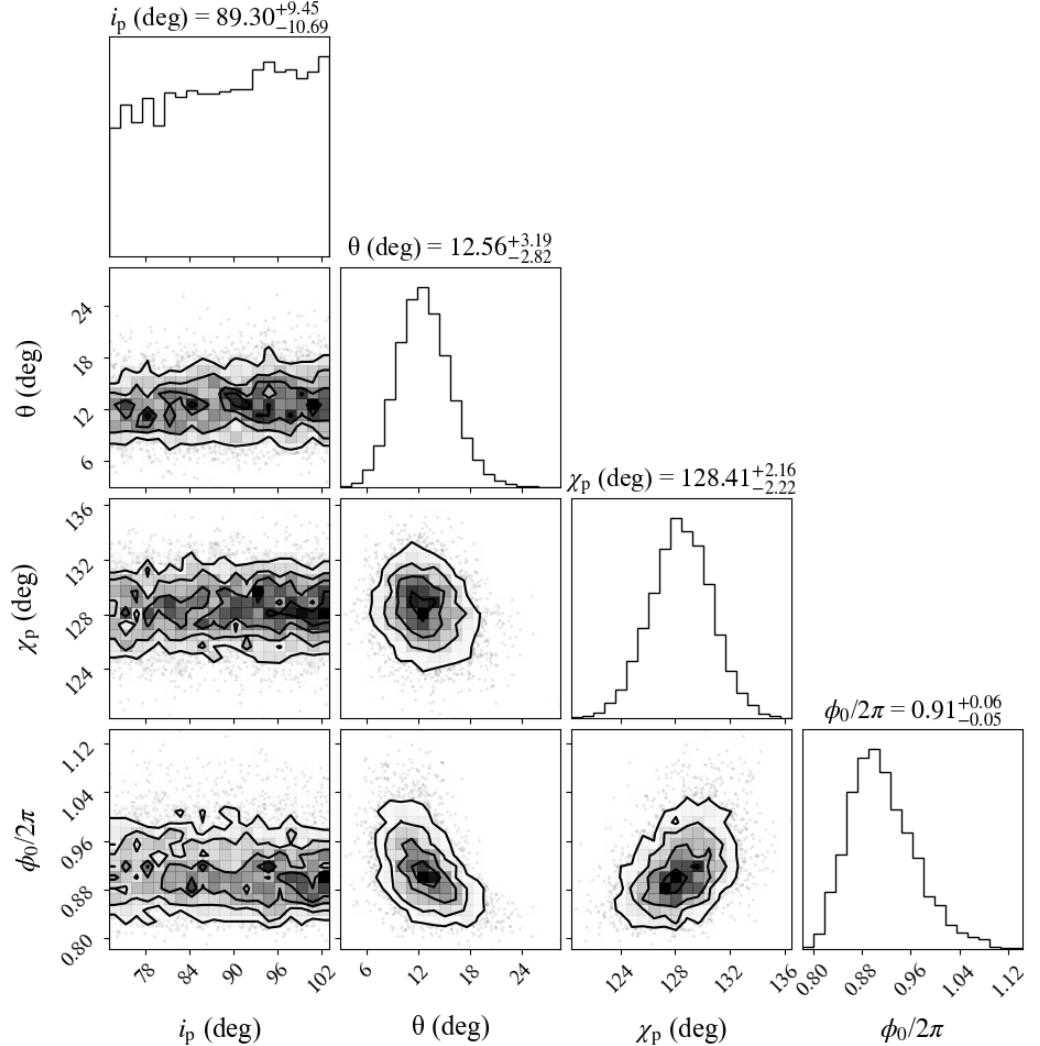}
\caption{Same as Figure \ref{mcmch}, but for the simultaneously fitting to phase-resolved \(\mathrm{PA_{1}}\) and \(\mathrm{PA_{2,3}}\)}
\label{mcmc1}
\end{figure*}

\clearpage
\bibliography{reference}

@ARTICLE{1976ApJ...206L..99M,
author = {{McClintock}, J. E. and {Rappaport}, S. and {Joss}, P. C. and {Bradt}, H. and {Buff}, J. and {Clark}, G. W.},
title = "{Discovery of a 283-second periodic variation in the X-ray source 3U 0900-40}",
journal = {The Astrophysical Journal},
year = 1976,
doi = {10.1086/182142}
}

@ARTICLE{1972ApJ...175L..19H,
author = {{Hiltner}, W. A. and {Werner}, J. and {Osmer}, P.},
title = "{Binary Nature of the B Supergiant in the Error Box of the VELA X-Ray Source}",
journal = {The Astrophysical Journal},
year=1972,
doi = {10.1086/180976}
}

@article{faucris.216777486,
 author = {Kreykenbohm, Ingo and Wilms, Jörn and Kretschmar, Peter and Torrejon, J. M. and Pottschmidt, K. and Hanke, Manfred and Santangelo, A. and Ferrigno, Carlo and Staubert, R.},
 doi = {10.1051/0004-6361:200809956},
 faupublication = {yes},
 journal = {Astronomy \& Astrophysics},
 pages = {511-525},
 peerreviewed = {Yes},
 title = {{High} variability in {Vela} {X}-1: giant flares and off states},
 volume = {492},
 year = {2008}
}

@article{Quaintrell2003TheMO,
  title={The mass of the neutron star in Vela X-1 and tidally induced non-radial oscillations in GP Vel},
  author={Hannah Quaintrell and A. J. Norton and T. D. C. Ash and P. Roche and B. Willems and Timothy R. Bedding and Ivan K. Baldry and Radboud University and Uk and Kildrummy Technologies Ltd and University of Glamorgan and University of Leicester and University of Sydney and Australia and John s Hopkins University and Usa and Anton Pannekoek Institute and The Netherlands},
  journal={Astronomy and Astrophysics},
  year={2003},
  volume={401},
  pages={313-323},
  doi={10.1051/0004-6361:20011122}
}

@article{GmenezGarca2016MeasuringTS,
  title={Measuring the stellar wind parameters in IGR J17544-2619 and Vela X-1 constrains the accretion physics in Supergiant Fast X-ray Transient and classical Supergiant X-ray Binaries},
  author={Angel G{\'i}menez-Garc{\'i}a and Tomer Shenar and J. M. Torrej{\'o}n and Lidia M. Oskinova and Silvia Mart{\'i}nez-N{\'u}{\~n}ez and Wolf-Rainer Hamann and Jose Joaqu{\'i}n Rodes-Roca and A. Gonz{\'a}lez-Gal{\'a}n and Javier Alonso-Santiago and Carlos Gonzalez-Fernandez and Guillermo Bernabeu and Andreas A C Sander},
  journal={arXiv: High Energy Astrophysical Phenomena},
  year={2016},
  doi ={10.1051/0004-6361/201527551 }
}

@article{Sato1986XrayPO,
  title={X-ray Probing of the Circumstellar Matter in the Vela X-1 System from Observations over an Eclipse Phase},
  author={N. Sato and Satio Hayakawa and Fumiaki Nagase and Kuniaki Masai and Tadayasu Dotani and Hajime Inoue and Fumiyoshi Makino and Kazuo Makishima and Takaya Ohashi},
  journal={Publications of the Astronomical Society of Japan},
  year={1986},
  volume={38},
  pages={731-750},
}

@article{Kretschmar2021RevisitingTA,
  title={Revisiting the archetypical wind accretor Vela X-1 in depth},
  author={Peter Kretschmar and Ileyk El Mellah and S. Mart'inez-N'unez and Felix Furst and Victoria Grinberg and Andreas A. C. Sander and Jakob van den Eijnden and Nathalie Degenaar and Jes'us Ma'iz-Apell'aniz and Francisco Jim'enez Esteban and M. Ramos-Lerate and E. Utrilla},
  journal={Astronomy and Astrophysics},
  year={2021},
  doi = {10.1051/0004-6361/202040272}
}

@article{Doroshenko2022DeterminationOX,
  title={Determination of X-ray pulsar geometry with IXPE polarimetry},
  author={Victor Doroshenko and Juri Poutanen and Sergey S. Tsygankov and Valery F. Suleimanov and Matteo Bachetti and Ilaria Caiazzo and Enrico Costa and Alessandro Di Marco and Jeremy S. Heyl and Fabio La Monaca and Fabio Muleri and Alexander A. Mushtukov and George G. Pavlov and Brian D. Ramsey and John Rankin and Andrea Santangelo and Paolo Soffitta and R{\"u}diger Staubert and Martin. C. Weisskopf and Silvia Zane and Iv{\'a}n Agudo and Lucio Angelo Antonelli and Luca Baldini and Wayne H. Baumgartner and Ronaldo Bellazzini and Stefano Bianchi and Stephen D. Bongiorno and Raffaella Bonino and Alessandro Brez and Niccolo’ Bucciantini and Fiamma Capitanio and Simone Castellano and Elisabetta Cavazzuti and Stefano Ciprini and Alessandra De Rosa and Ettore Del Monte and Laura Di Gesu and Niccolo’ Di Lalla and Immacolata Donnarumma and Michal Dovc̆iak and Steven R. Ehlert and Teruaki Enoto and Yuri Evangelista and Sergio Fabiani and Riccardo Ferrazzoli and Javier A. Garc{\'i}a and Shuichi Gunji and Kiyoshi Hayashida and Wataru Buz Iwakiri and Svetlana G. Jorstad and Vladim{\'i}r Karas and Takao Kitaguchi and Jeffery J. Kolodziejczak and Henric Krawczynski and Luca Latronico and Ioannis Liodakis and Simone Maldera and Alberto Manfreda and Fr{\'e}d{\'e}ric Marin and Andrea Marinucci and Alan P. Marscher and Herman L. Marshall and Giorgio Matt and Ikuyuki Mitsuishi and Tsunefumi Mizuno and C-Y. Ng and Stephen L. O’Dell and Nicola Omodei and Chiara Oppedisano and Alessandro Papitto and Abel Lawrence Peirson and Matteo Perri and Melissa Pesce-Rollins and Maura Pilia and Andrea Possenti and Simonetta Puccetti and Ajay Ratheesh and Roger W. Romani and Carmelo Sgro’ and Patrick O. Slane and Gloria Spandre and Rashid A. Sunyaev and Toru Tamagawa and Fabrizio Tavecchio and Roberto Taverna and Yuzuru Tawara and Allyn F. Tennant and Nicolas Thomas and Francesco Tombesi and Alessio Trois and Roberto Turolla and Jacco Vink and Kinwah Wu and Fei Xie},
  journal={Nature Astronomy},
  year={2022},
  volume={6},
  pages={1433 - 1443},
  doi ={10.1038/s41550-022-01799-5}
}

@article{Heyl2024ComplexRD,
  title={Complex rotational dynamics of the neutron star in Hercules X-1 revealed by X-ray polarization},
  author={Jeremy S. Heyl and Victor Doroshenko and Denis Gonz{\'a}lez-Caniulef and Ilaria Caiazzo and Juri Poutanen and Alexander A. Mushtukov and Sergey S. Tsygankov and Demet Kırmızıbayrak and Matteo Bachetti and George G. Pavlov and Sofia V. Forsblom and Christian Malacaria and Valery F. Suleimanov and Iv{\'a}n Agudo and Lucio Angelo Antonelli and Luca Baldini and Wayne H. Baumgartner and Ronaldo Bellazzini and Stefano Bianchi and Stephen D. Bongiorno and Raffaella Bonino and Alessandro Brez and Niccol{\'o} Bucciantini and Fiamma Capitanio and Simone Castellano and Elisabetta Cavazzuti and Chien-Ting J. Chen and Stefano Ciprini and Enrico Costa and Alessandra De Rosa and Ettore Del Monte and Laura Di Gesu and Niccolo’ Di Lalla and Alessandro Di Marco and Immacolata Donnarumma and Michal Dovc̆iak and Steven R. Ehlert and Teruaki Enoto and Yuri Evangelista and Sergio Fabiani and Riccardo Ferrazzoli and Javier A. Garc{\'i}a and Shuichi Gunji and Kiyoshi Hayashida and Wataru Buz Iwakiri and Svetlana G. Jorstad and Philip Kaaret and Vladim{\'i}r Karas and Fabian Kislat and Takao Kitaguchi and Jeffery J. Kolodziejczak and Henric Krawczynski and Fabio La Monaca and Luca Latronico and Ioannis Liodakis and Simone Maldera and Alberto Manfreda and Fr{\'e}d{\'e}ric Marin and Andrea Marinucci and Alan P. Marscher and Herman L. Marshall and Francesco Massaro and Giorgio Matt and Ikuyuki Mitsuishi and Tsunefumi Mizuno and Fabio Muleri and Michela Negro and Chi-Yung Ng and Stephen L. O’Dell and Nicola Omodei and Chiara Oppedisano and Alessandro Papitto and Abel Lawrence Peirson and Matteo Perri and Melissa Pesce-Rollins and Pierre-Olivier Petrucci and Maura Pilia and Andrea Possenti and Simonetta Puccetti and Brian D. Ramsey and John Rankin and Ajay Ratheesh and Oliver J. Roberts and Roger W. Romani and Carmelo Sgro’ and Patrick O. Slane and Paolo Soffitta and Gloria Spandre and Douglas A. Swartz and Toru Tamagawa and Fabrizio Tavecchio and Roberto Taverna and Yuzuru Tawara and Allyn F. Tennant and Nicholas E. Thomas and Francesco Tombesi and Alessio Trois and Roberto Turolla and Jacco Vink and Martin. C. Weisskopf and Kinwah Wu and Fei Xie and Silvia Zane},
  journal={Nature Astronomy},
  year={2024},
  doi={10.1038/s41550-024-02295-8}
}

@article{Zhao2024PolarizationPO,
       author = {{Zhao}, Q.~C. and {Li}, H.~C. and {Tao}, L. and {Feng}, H. and {Zhang}, S.~N. and {Walter}, R. and {Ge}, M.~Y. and {Tong}, H. and {Ji}, L. and {Zhang}, L. and {Qu}, J.~L. and {Huang}, Y. and {Ma}, X. and {Zhang}, S. and {Yin}, Q.~Q. and {Yin}, H.~X. and {Ma}, R.~C. and {Zhao}, S.~J. and {Li}, P.~P. and {Yang}, Z.~X. and {Liu}, H.~X. and {Yu}, W. and {Huang}, Y.~M. and {Li}, Z.~X. and {Li}, Y.~J. and {Xiao}, J.~Y. and {Zhao}, K.},
        title = "{Polarization perspectives on Hercules X-1: further constraining the geometry}",
      journal = {Monthly Notices of the Royal Astronomical Society},
         year = 2024,
        month = jul,
       volume = {531},
       number = {4},
        pages = {3935-3949},
          doi = {10.1093/mnras/stae1173},
archivePrefix = {arXiv},
       eprint = {2405.00509},
 primaryClass = {astro-ph.HE},
       adsurl = {https://ui.adsabs.harvard.edu/abs/2024MNRAS.531.3935Z}
}

@article{Tsygankov2022TheXP,
  title={The X-Ray Polarimetry View of the Accreting Pulsar Cen X-3},
  author={Sergey S. Tsygankov and Victor Doroshenko and Juri Poutanen and Jeremy S. Heyl and Alexander A. Mushtukov and Ilaria Caiazzo and Alessandro Di Marco and Sofia V. Forsblom and Denis Gonz{\'a}lez-Caniulef and Moritz Klawin and Fabio La Monaca and Christian Malacaria and Herman L. Marshall and Fabio Muleri and Mason Ng and Valery F. Suleimanov and Rashid A. Sunyaev and Roberto Turolla and Iv{\'a}n Agudo and Lucio Angelo Antonelli and Matteo Bachetti and Luca Baldini and Wayne H. Baumgartner and Ronaldo Bellazzini and Stefano Bianchi and Stephen D. Bongiorno and Raffaella Bonino and Alessandro Brez and Niccolo’ Bucciantini and Fiamma Capitanio and Simone Castellano and Elisabetta Cavazzuti and Stefano Ciprini and Enrico Costa and Alessandra De Rosa and Ettore Del Monte and Laura Di Gesu and Niccol{\'o} Di Lalla and Immacolata Donnarumma and Michal Dov\v{c}iak and Steven R. Ehlert and Teruaki Enoto and Yuri Evangelista and Sergio Fabiani and Riccardo Ferrazzoli and Javier A. Garc{\'i}a and Shuichi Gunji and Kiyoshi Hayashida and Wataru Buz Iwakiri and Svetlana G. Jorstad and Vladim{\'i}r Karas and Takao Kitaguchi and Jeffery J. Kolodziejczak and Henric Krawczynski and Luca Latronico and Ioannis Liodakis and Simone Maldera and Alberto Manfreda and Fr{\'e}d{\'e}ric Marin and Andrea Marinucci and Alan P. Marscher and Giorgio Matt and Ikuyuki Mitsuishi and Tsunefumi Mizuno and C-Y. Ng and Stephen L. O’Dell and Nicola Omodei and Chiara Oppedisano and Alessandro Papitto and George G. Pavlov and Abel Lawrence Peirson and Matteo Perri and Melissa Pesce-Rollins and Pierre-Olivier Petrucci and Maura Pilia and Andrea Possenti and Simonetta Puccetti and Brian D. Ramsey and John Rankin and Ajay Ratheesh and Roger W. Romani and Carmelo Sgro’ and Patrick O. Slane and Paolo Soffitta and Gloria Spandre and Toru Tamagawa and Fabrizio Tavecchio and Roberto Taverna and Yuzuru Tawara and Allyn F. Tennant and Nicholas E. Thomas and Francesco Tombesi and Alessio Trois and Jacco Vink and Martin. C. Weisskopf and Kinwah Wu and Fei Xie and Silvia Zane},
  journal={The Astrophysical Journal Letters},
  year={2022},
  volume={941},
  doi={10.3847/2041-8213/aca486}
}

@article{Tsygankov2023XrayPG,
  title={X-ray pulsar GRO~J1008-57 as an orthogonal rotator},
  author={Sergey S. Tsygankov and Victor Doroshenko and Alexander A. Mushtukov and Juri Poutanen and Alessandro Di Marco and Jeremy S. Heyl and Fabio La Monaca and Sofia V. Forsblom and Christian Malacaria and Herman L. Marshall and Valery F. Suleimanov and Jiř{\'i} Svoboda and Roberto Taverna and Francesco Ursini and Iv{\'a}n Agudo and Lucio Angelo Antonelli and Matteo Bachetti and Luca Baldini and Wayne H. Baumgartner and Ronaldo Bellazzini and Stefano Bianchi and Stephen D. Bongiorno and Raffaella Bonino and Alessandro Brez and Niccolo’ Bucciantini and Fiamma Capitanio and Simone Castellano and Elisabetta Cavazzuti and Chien-Ting J. Chen and Stefano Ciprini and Enrico Costa and Alessandra De Rosa and Ettore Del Monte and Laura Di Gesu and Niccol{\'o} Di Lalla and Immacolata Donnarumma and Michal Dovvciak and Steven R. Ehlert and Teruaki Enoto and Yuri Evangelista and Sergio Fabiani and Riccardo Ferrazzoli and Javier A. Garc{\'i}a and Shuichi Gunji and Kiyoshi Hayashida and Wataru Buz Iwakiri and Svetlana G. Jorstad and Philip E. Kaaret and Vladim{\'i}r Karas and Fabian Kislat and Takao Kitaguchi and Jeffery J. Kolodziejczak and Henric Krawczynski and Luca Latronico and Ioannis Liodakis and Simone Maldera and Alberto Manfreda and Fr{\'e}d{\'e}ric Marin and Andrea Marinucci and Alan P. Marscher and Francesco Massaro and Giorgio Matt and Ikuyuki Mitsuishi and Tsunefumi Mizuno and Fabio Muleri and Michela Negro and C-Y. Ng and Stephen L. O’Dell and Nicola Omodei and Chiara Oppedisano and Alessandro Papitto and George G. Pavlov and Abel Lawrence Peirson and Matteo Perri and Melissa Pesce-Rollins and Pierre-Olivier Petrucci and Maura Pilia and Andrea Possenti and Simonetta Puccetti and Brian D. Ramsey and John Rankin and Ajay Ratheesh and Oliver J. Roberts and Roger W. Romani and Carmelo Sgro’ and Patrick O. Slane and Paolo Soffitta and Gloria Spandre and Douglas A. Swartz and Toru Tamagawa and Fabrizio Tavecchio and Yuzuru Tawara and Allyn F. Tennant and Nicholas E. Thomas and Francesco Tombesi and Alessio Trois and Roberto Turolla and Jacco Vink and Martin. C. Weisskopf and Kinwah Wu and Fei Xie and Silvia Zane},
  journal={Astronomy \& Astrophysics},
  year={2023},
  doi = {10.1051/0004-6361/202346134}
}

@article{Marshall2022ObservationsO4,
       author = {{Marshall}, Herman L. and {Ng}, Mason and {Rogantini}, Daniele and {Heyl}, Jeremy and {Tsygankov}, Sergey S. and {Poutanen}, Juri and {Costa}, Enrico and {Zane}, Silvia and {Malacaria}, Christian and {Agudo}, Iv{\'a}n and {Antonelli}, Lucio A. and {Bachetti}, Matteo and {Baldini}, Luca and {Baumgartner}, Wayne H. and {Bellazzini}, Ronaldo and {Bianchi}, Stefano and {Bongiorno}, Stephen D. and {Bonino}, Raffaella and {Brez}, Alessandro and {Bucciantini}, Niccol{\`o} and {Capitanio}, Fiamma and {Castellano}, Simone and {Cavazzuti}, Elisabetta and {Ciprini}, Stefano and {De Rosa}, Alessandra and {Del Monte}, Ettore and {Di Gesu}, Laura and {Di Lalla}, Niccol{\`o} and {Di Marco}, Alessandro and {Donnarumma}, Immacolata and {Doroshenko}, Victor and {Dov{\v{c}}iak}, Michal and {Ehlert}, Steven R. and {Enoto}, Teruaki and {Evangelista}, Yuri and {Fabiani}, Sergio and {Ferrazzoli}, Riccardo and {Garcia}, Javier A. and {Gunji}, Shuichi and {Hayashida}, Kiyoshi and {Iwakiri}, Wataru and {Jorstad}, Svetlana G. and {Karas}, Vladimir and {Kitaguchi}, Takao and {Kolodziejczak}, Jeffery J. and {Krawczynski}, Henric and {La Monaca}, Fabio and {Latronico}, Luca and {Liodakis}, Ioannis and {Maldera}, Simone and {Manfreda}, Alberto and {Marin}, Fr{\'e}d{\'e}ric and {Marinucci}, Andrea and {Marscher}, Alan P. and {Matt}, Giorgio and {Mitsuishi}, Ikuyuki and {Mizuno}, Tsunefumi and {Muleri}, Fabio and {Ng}, C. -Y. and {O'Dell}, Stephen L. and {Omodei}, Nicola and {Oppedisano}, Chiara and {Papitto}, Alessandro and {Pavlov}, George G. and {Peirson}, Abel L. and {Perri}, Matteo and {Pesce-Rollins}, Melissa and {Petrucci}, Pierre-Olivier and {Pilia}, Maura and {Possenti}, Andrea and {Puccetti}, Simonetta and {Ramsey}, Brian D. and {Rankin}, John and {Ratheesh}, Ajay and {Romani}, Roger W. and {Sgr{\`o}}, Carmelo and {Slane}, Patrick and {Soffitta}, Paolo and {Spandre}, Gloria and {Tamagawa}, Toru and {Tavecchio}, Fabrizio and {Taverna}, Roberto and {Tawara}, Yuzuru and {Tennant}, Allyn F. and {Thomas}, Nicholas E. and {Tombesi}, Francesco and {Trois}, Alessio and {Turolla}, Roberto and {Vink}, Jacco and {Weisskopf}, Martin C. and {Wu}, Kinwah and {Xie}, Fei and {IXPE Collaboration} and {Schulz}, Norbert S. and {Chakrabarty}, Deepto},
        title = "{Observations of 4U 1626-67 with the Imaging X-Ray Polarimetry Explorer}",
      journal = {The Astrophysical Journal},
         year = 2022,
        month = nov,
       volume = {940},
       number = {1},
          eid = {70},
        pages = {70},
          doi = {10.3847/1538-4357/ac98c2},
archivePrefix = {arXiv},
       eprint = {2210.03194},
 primaryClass = {astro-ph.HE},
       adsurl = {https://ui.adsabs.harvard.edu/abs/2022ApJ...940...70M}
}

@article{Mushtukov2023XrayPO,
  title={X-ray polarimetry of X-ray pulsar X Persei: another orthogonal rotator?},
  author={Alexander A. Mushtukov and Sergey S. Tsygankov and Juri Poutanen and Victor Doroshenko and Alexander Salganik and E. Costa and Alessandro Di Marco and Jeremy S. Heyl and Fabio La Monaca and Alexander A. Lutovinov and I. A. Mereminsky and Alessandro Papitto and Andrei N. Semena and A Shtykovsky and Valery F. Suleimanov and Sofia V. Forsblom and Denis Gonz{\'a}lez-Caniulef and Christian Malacaria and Rashid A. Sunyaev and Iv{\'a}n Agudo and Lucio Angelo Antonelli and Matteo Bachetti and Luca Baldini and Wayne H. Baumgartner and Ronaldo Bellazzini and S Bianchi and Stephen D. Bongiorno and Raffaella Bonino and Alessandro Brez and Niccolo’ Bucciantini and Fiamma Capitanio and Simone Castellano and Elisabetta Cavazzuti and C.T. Chen and Stefano Ciprini and Alessandra De Rosa and Ettore Del Monte and Laura Di Gesu and Niccolo’ Di Lalla and Immacolata Donnarumma and Michal Dovc̆iak and Steven R. Ehlert and Teruaki Enoto and Yuri Evangelista and Sergio Fabiani and Riccardo Ferrazzoli and J. A. Garc{\'i}a and S. Gunji and Kiyoshi Hayashida and Wataru Buz Iwakiri and Svetlana G. Jorstad and P. Kaaret and Vladim{\'i}r Karas and Fabian Kislat and Takao Kitaguchi and Jeffery J. Kolodziejczak and Henric Krawczynski and Luca Latronico and Ioannis Liodakis and Simone Maldera and Alberto Manfreda and Fr{\'e}d{\'e}ric Marin and Alan P. Marscher and Herman L. Marshall and Francesco Massaro and G. Matt and Ikuyuki Mitsuishi and Tsunefumi Mizuno and Fabio Muleri and Michela Negro and C-Y. Ng and Stephen L. O’Dell and Nicola Omodei and Chiara Oppedisano and George G. Pavlov and Abel Lawrence Peirson and Matteo Perri and Melissa Pesce-Rollins and Pierre-Olivier Petrucci and Maura Pilia and Andrea Possenti and Simonetta Puccetti and Brian D. Ramsey and John Rankin and Ajay Ratheesh and Oliver J. Roberts and R. W. Romani and Carmelo Sgro’ and Patrick O. Slane and Paolo Soffitta and Gloria Spandre and Douglas A. Swartz and Toru Tamagawa and Fabrizio Tavecchio and Roberto Taverna and Y. Tawara and Allyn F. Tennant and Nicholas E. Thomas and Francesco Tombesi and Alessio Trois and Roberto Turolla and Jacco Vink and Martin. C. Weisskopf and K Wu and Fei Xie and Silvia Zane},
  journal={Monthly Notices of the Royal Astronomical Society},
  year={2023},
  doi={10.1093/mnras/stad1961}
}

@article{Forsblom2023IXPEOO,
       author = {{Forsblom}, Sofia V. and {Poutanen}, Juri and {Tsygankov}, Sergey S. and {Bachetti}, Matteo and {Di Marco}, Alessandro and {Doroshenko}, Victor and {Heyl}, Jeremy and {La Monaca}, Fabio and {Malacaria}, Christian and {Marshall}, Herman L. and {Muleri}, Fabio and {Mushtukov}, Alexander A. and {Pilia}, Maura and {Rogantini}, Daniele and {Suleimanov}, Valery F. and {Taverna}, Roberto and {Xie}, Fei and {Agudo}, Iv{\'a}n and {Antonelli}, Lucio A. and {Baldini}, Luca and {Baumgartner}, Wayne H. and {Bellazzini}, Ronaldo and {Bianchi}, Stefano and {Bongiorno}, Stephen D. and {Bonino}, Raffaella and {Brez}, Alessandro and {Bucciantini}, Niccol{\`o} and {Capitanio}, Fiamma and {Castellano}, Simone and {Cavazzuti}, Elisabetta and {Chen}, Chien-Ting and {Ciprini}, Stefano and {Costa}, Enrico and {De Rosa}, Alessandra and {Del Monte}, Ettore and {Di Gesu}, Laura and {Di Lalla}, Niccol{\`o} and {Donnarumma}, Immacolata and {Dov{\v{c}}iak}, Michal and {Ehlert}, Steven R. and {Enoto}, Teruaki and {Evangelista}, Yuri and {Fabiani}, Sergio and {Ferrazzoli}, Riccardo and {Garcia}, Javier A. and {Gunji}, Shuichi and {Hayashida}, Kiyoshi and {Iwakiri}, Wataru and {Jorstad}, Svetlana G. and {Kaaret}, Philip and {Karas}, Vladimir and {Kitaguchi}, Takao and {Kolodziejczak}, Jeffery J. and {Krawczynski}, Henric and {Latronico}, Luca and {Liodakis}, Ioannis and {Maldera}, Simone and {Manfreda}, Alberto and {Marin}, Fr{\'e}d{\'e}ric and {Marinucci}, Andrea and {Marscher}, Alan P. and {Matt}, Giorgio and {Mitsuishi}, Ikuyuki and {Mizuno}, Tsunefumi and {Negro}, Michela and {Ng}, Chi-Yung and {O'Dell}, Stephen L. and {Omodei}, Nicola and {Oppedisano}, Chiara and {Papitto}, Alessandro and {Pavlov}, George G. and {Peirson}, Abel L. and {Perri}, Matteo and {Pesce-Rollins}, Melissa and {Petrucci}, Pierre-Olivier and {Possenti}, Andrea and {Puccetti}, Simonetta and {Ramsey}, Brian D. and {Rankin}, John and {Ratheesh}, Ajay and {Roberts}, Oliver J. and {Romani}, Roger W. and {Sgr{\`o}}, Carmelo and {Slane}, Patrick and {Soffitta}, Paolo and {Spandre}, Gloria and {Sunyaev}, Rashid A. and {Swartz}, Douglas A. and {Tamagawa}, Toru and {Tavecchio}, Fabrizio and {Tawara}, Yuzuru and {Tennant}, Allyn F. and {Thomas}, Nicholas E. and {Tombesi}, Francesco and {Trois}, Alessio and {Turolla}, Roberto and {Vink}, Jacco and {Weisskopf}, Martin C. and {Wu}, Kinwah and {Zane}, Silvia and {IXPE Collaboration}},
        title = "{IXPE Observations of the Quintessential Wind-accreting X-Ray Pulsar Vela X-1}",
      journal = {The Astrophysical Journal Letters},
         year = 2023,
        month = apr,
       volume = {947},
       number = {2},
          eid = {L20},
        pages = {L20},
          doi = {10.3847/2041-8213/acc391},
archivePrefix = {arXiv},
       eprint = {2303.01800},
 primaryClass = {astro-ph.HE},
       adsurl = {https://ui.adsabs.harvard.edu/abs/2023ApJ...947L..20F}
}

@article{Malacaria2023APO,
  author = {{Malacaria}, Christian and {Heyl}, Jeremy and {Doroshenko}, Victor and {Tsygankov}, Sergey S. and {Poutanen}, Juri and {Forsblom}, Sofia V. and {Capitanio}, Fiamma and {Di Marco}, Alessandro and {Du}, Yujia and {Ducci}, Lorenzo and {La Monaca}, Fabio and {Lutovinov}, Alexander A. and {Marshall}, Herman L. and {Mereminskiy}, Ilya A. and {Molkov}, Sergey V. and {Mushtukov}, Alexander A. and {Ng}, Mason and {Petrucci}, Pierre-Olivier and {Santangelo}, Andrea and {Shtykovsky}, Andrey E. and {Suleimanov}, Valery F. and {Agudo}, Iv{\'a}n and {Antonelli}, Lucio A. and {Bachetti}, Matteo and {Baldini}, Luca and {Baumgartner}, Wayne H. and {Bellazzini}, Ronaldo and {Bianchi}, Stefano and {Bongiorno}, Stephen D. and {Bonino}, Raffaella and {Brez}, Alessandro and {Bucciantini}, Niccol{\`o} and {Castellano}, Simone and {Cavazzuti}, Elisabetta and {Chen}, Chien-Ting and {Ciprini}, Stefano and {Costa}, Enrico and {De Rosa}, Alessandra and {Del Monte}, Ettore and {Di Gesu}, Laura and {Di Lalla}, Niccol{\`o} and {Donnarumma}, Immacolata and {Dov{\v{c}}iak}, Michal and {Ehlert}, Steven R. and {Enoto}, Teruaki and {Evangelista}, Yuri and {Fabiani}, Sergio and {Ferrazzoli}, Riccardo and {Garcia}, Javier A. and {Gunji}, Shuichi and {Hayashida}, Kiyoshi and {Iwakiri}, Wataru and {Jorstad}, Svetlana G. and {Kaaret}, Philip and {Karas}, Vladimir and {Kislat}, Fabian and {Kitaguchi}, Takao and {Kolodziejczak}, Jeffery J. and {Krawczynski}, Henric and {Latronico}, Luca and {Liodakis}, Ioannis and {Maldera}, Simone and {Manfreda}, Alberto and {Marin}, Fr{\'e}d{\'e}ric and {Marinucci}, Andrea and {Marscher}, Alan P. and {Massaro}, Francesco and {Matt}, Giorgio and {Mitsuishi}, Ikuyuki and {Mizuno}, Tsunefumi and {Muleri}, Fabio and {Negro}, Michela and {Ng}, Chi-Yung and {O'Dell}, Stephen L. and {Omodei}, Nicola and {Oppedisano}, Chiara and {Papitto}, Alessandro and {Pavlov}, George G. and {Peirson}, Abel L. and {Perri}, Matteo and {Pesce-Rollins}, Melissa and {Pilia}, Maura and {Possenti}, Andrea and {Puccetti}, Simonetta and {Ramsey}, Brian D. and {Rankin}, John and {Ratheesh}, Ajay and {Roberts}, Oliver J. and {Romani}, Roger W. and {Sgr{\`o}}, Carmelo and {Slane}, Patrick and {Soffitta}, Paolo and {Spandre}, Gloria and {Swartz}, Douglas A. and {Tamagawa}, Toru and {Tavecchio}, Fabrizio and {Taverna}, Roberto and {Tawara}, Yuzuru and {Tennant}, Allyn F. and {Thomas}, Nicholas E. and {Tombesi}, Francesco and {Trois}, Alessio and {Turolla}, Roberto and {Vink}, Jacco and {Weisskopf}, Martin C. and {Wu}, Kinwah and {Xie}, Fei and {Zane}, Silvia},
        title = "{A polarimetrically oriented X-ray stare at the accreting pulsar EXO 2030+375}",
      journal = {Astronomy \& Astrophysics},
         year = 2023,
        month = jul,
       volume = {675},
          eid = {A29},
        pages = {A29},
          doi = {10.1051/0004-6361/202346581},
archivePrefix = {arXiv},
       eprint = {2304.00925},
 primaryClass = {astro-ph.HE},
       adsurl = {https://ui.adsabs.harvard.edu/abs/2023A&A...675A..29M}
}

@article{Suleimanov2023XrayPO,
  title={X-ray polarimetry of the accreting pulsar GX 301−2},
  author={Valery F. Suleimanov and Sofia V. Forsblom and Sergey S. Tsygankov and Juri Poutanen and Victor Doroshenko and Rosalia Doroshenko and Fiamma Capitanio and Alessandro Di Marco and Denis Gonz'alez-Caniulef and Jeremy S. Heyl and Fabio La Monaca and Alexander A. Lutovinov and Sergey V. Molkov and Christian Malacaria and Alexander A. Mushtukov and A Shtykovsky and Iv{\'a}n Agudo and Lucio Angelo Antonelli and Matteo Bachetti and Luca Baldini and Wayne H. Baumgartner and Ronaldo Bellazzini and Stefano Bianchi and Stephen D. Bongiorno and Raffaella Bonino and Alessandro Brez and Niccolo’ Bucciantini and Simone Castellano and Elisabetta Cavazzuti and Chien-Ting J. Chen and Stefano Ciprini and Enrico Costa and Alessandra De Rosa and Ettore Del Monte and Laura Di Gesu and Gímenez-Garcíarazzoli and Javier A. Garc{\'i}a and Shuichi Gunji and Kiyoshi Hayashida and Wataru Buz Iwakiri and Svetlana G. Jorstad and Philip E. Kaaret and Vladim{\'i}r Karas and Fabian Kislat and Takao Kitaguchi and Jeffery J. Kolodziejczak and Henric Krawczynski and Luca Latronico and Ioannis Liodakis and Simone Maldera and Alberto Manfreda and Fr{\'e}d{\'e}ric Marin and Andrea Marinucci and Alan P. Marscher and Herman L. Marshall and Francesco Massaro and Giorgio Matt and Ikuyuki Mitsuishi and Tsunefumi Mizuno and Fabio Muleri and Michela Negro and C-Y. Ng and Stephen L. O’Dell and Nicola Omodei and Chiara Oppedisano and Alessandro Papitto and George G. Pavlov and Abel Lawrence Peirson and Matteo Perri and Melissa Pesce Rollins and Pierre-Olivier Petrucci and Maura Pilia and Andrea Possenti and Simonetta Puccetti and Brian D. Ramsey and John Rankin and Ajay Ratheesh and Oliver J. Roberts and Roger W. Romani and Carmelo Sgro’ and Patrick O. Slane and Paolo Soffitta and Gloria Spandre and Douglas A. Swartz and Toru Tamagawa and Fabrizio Tavecchio and Roberto Taverna and Yuzuru Tawara and Allyn F. Tennant and Nicholas E. Thomas and Francesco Tombesi and Alessio Trois and Roberto Turolla and Jacco Vink and Martin. C. Weisskopf and Kinwah Wu and Fei Xie and Silvia Zane},
  journal={Astronomy \& Astrophysics},
  year={2023},
  doi={10.1051/0004-6361/202346994}
}

@article{Doroshenko2023ComplexVI,
  title={Complex variations in X-ray polarization in the X-ray pulsar LS V +44 17/RX J0440.9+4431},
  author={Victor Doroshenko and Juri Poutanen and Jeremy S. Heyl and Sergey S. Tsygankov and Ilaria Caiazzo and Roberto Turolla and Alexandra Veledina and Martin. C. Weisskopf and Sofia V. Forsblom and Denis Gonz{\'a}lez-Caniulef and Vladislav Loktev and Christian Malacaria and Alexander A. Mushtukov and Valery F. Suleimanov and Alexander A. Lutovinov and Ilya A Mereminskiy and Sergey V. Molkov and Alexander Salganik and Andrea Santangelo and Andrei V. Berdyugin and Vadim Kravtsov and Anagha P. Nitindala and Iv{\'a}n Agudo and Lucio Angelo Antonelli and Matteo Bachetti and Luca Baldini and Wayne H. Baumgartner and Ronaldo Bellazzini and Stefano Bianchi and Stephen D. Bongiorno and Raffaella Bonino and Alessandro Brez and Niccolo’ Bucciantini and Fiamma Capitanio and Simone Castellano and Elisabetta Cavazzuti and Chien-Ting J. Chen and Stefano Ciprini and Enrico Costa and Alessandra De Rosa and Ettore Del Monte and Laura Di Gesu and Niccol{\'o} Di Lalla and Alessandro Di Marco and Immacolata Donnarumma and Michal Dovvciak and Steven R. Ehlert and Teruaki Enoto and Yuri Evangelista and Sergio Fabiani and Riccardo Ferrazzoli and Javier A. Garc{\'i}a and Shuichi Gunji and Kiyoshi Hayashida and Wataru Buz Iwakiri and Svetlana G. Jorstad and Philip E. Kaaret and Vladim{\'i}r Karas and Fabian Kislat and Takao Kitaguchi and Jeffery J. Kolodziejczak and Henric Krawczynski and Fabio La Monaca and Luca Latronico and Ioannis Liodakis and Simone Maldera and Alberto Manfreda and Fr{\'e}d{\'e}ric Marin and Andrea Marinucci and Alan P. Marscher and Herman L. Marshall and Francesco Massaro and Giorgio Matt and Ikuyuki Mitsuishi and Tsunefumi Mizuno and Fabio Muleri and Michela Negro and C-Y. Ng and Stephen L. O’Dell and Nicola Omodei and Chiara Oppedisano and Alessandro Papitto and George G. Pavlov and Abel Lawrence Peirson and Matteo Perri and Melissa Pesce-Rollins and Pierre-Olivier Petrucci and Maura Pilia and Andrea Possenti and Simonetta Puccetti and Brian D. Ramsey and John Rankin and Ajay Ratheesh and Oliver J. Roberts and Roger W. Romani and Carmelo Sgro’ and Patrick O. Slane and Paolo Soffitta and Gloria Spandre and Douglas A. Swartz and Toru Tamagawa and Fabrizio Tavecchio and Roberto Taverna and Yuzuru Tawara and Allyn F. Tennant and Nicholas E. Thomas and Francesco Tombesi and Alessio Trois and Jacco Vink and Kinwah Wu and Fei Xie and Silvia Zane},
  journal={Astronomy \& Astrophysics},
  year={2023},
  doi={10.1051/0004-6361/202347088}
}

@article{Poutanen2024StudyingGO,
  title={Studying geometry of the ultraluminous X-ray pulsar Swift J0243.6+6124 using X-ray and optical polarimetry},
  author={Juri Poutanen and Sergey S. Tsygankov and Victor Doroshenko and Sofia V. Forsblom and Peter Jenke and Philip Kaaret and Andrei V. Berdyugin and Dmitry Blinov and Vadim Kravtsov and Ioannis Liodakis and Anastasia Tzouvanou and Alessandro Di Marco and Jeremy S. Heyl and Fabio La Monaca and Alexander A. Mushtukov and George G. Pavlov and Alexander Salganik and Alexandra Veledina and Martin. C. Weisskopf and Silvia Zane and Vladislav Loktev and Valery F. Suleimanov and Colleen A. Wilson-Hodge and Svetlana V. Berdyugina and Masato Kagitani and Vilppu Piirola and Takeshi Sakanoi and Iv{\'a}n Agudo and Lucio Angelo Antonelli and Matteo Bachetti and Luca Baldini and Wayne H. Baumgartner and Ronaldo Bellazzini and Stefano Bianchi and Stephen D. Bongiorno and Raffaella Bonino and Alessandro Brez and Niccol{\'o} Bucciantini and Fiamma Capitanio and Simone Castellano and Elisabetta Cavazzuti and Chien-Ting J. Chen and Stefano Ciprini and Enrico Costa and Alessandra De Rosa and Ettore Del Monte and Laura Di Gesu and Niccol{\'o} Di Lalla and Immacolata Donnarumma and Michal Dovc̆iak and Steven R. Ehlert and Teruaki Enoto and Yuri Evangelista and Sergio Fabiani and Riccardo Ferrazzoli and Javier A. Garc{\'i}a and Shuichi Gunji and Kiyoshi Hayashida and Wataru Buz Iwakiri and Svetlana G. Jorstad and Vladimir Karas and Fabian Kislat and Takao Kitaguchi and Jeffery J. Kolodziejczak and Luca Latronico and Simone Maldera and Alberto Manfreda and Fr{\'e}d{\'e}ric Marin and Andrea Marinucci and Alan P. Marscher and Herman L. Marshall and Francesco Massaro and Giorgio Matt and Ikuyuki Mitsuishi and Tsunefumi Mizuno and Fabio Muleri and Michela Negro and Chi-Yung Ng and Stephen L. O’Dell and Nicola Omodei and Chiara Oppedisano and Alessandro Papitto and Abel Lawrence Peirson and Matteo Perri and Melissa Pesce-Rollins and Pierre-Olivier Petrucci and Maura Pilia and Andrea Possenti and Simonetta Puccetti and Brian D. Ramsey and John Rankin and Ajay Ratheesh and Oliver J. Roberts and Roger W. Romani and Carmelo Sgro’ and Patrick O. Slane and Paolo Soffitta and Gloria Spandre and Douglas A. Swartz and Toru Tamagawa and Fabrizio Tavecchio and Roberto Taverna and Yuzuru Tawara and Allyn F. Tennant and Nicholas E. Thomas and Francesco Tombesi and Alessio Trois and Roberto Turolla and Jacco Vink and Kinwah Wu and Fei Xie},
  journal={Astronomy \& Astrophysics},
  year={2024},
  doi={10.1051/0004-6361/202450696}
}

@article{Forsblom2024ProbingTP,
  title={Probing the polarized emission from SMC X-1: The brightest X-ray pulsar observed by IXPE},
  author={Sofia V. Forsblom and Sergey S. Tsygankov and Juri Poutanen and Victor Doroshenko and Alexander A. Mushtukov and Mason Ng and Swati Ravi and Herman L. Marshall and Alessandro Di Marco and Fabio La Monaca and Christian Malacaria and Guglielmo Mastroserio and Vladislav Loktev and Andrea Possenti and Valery F. Suleimanov and Roberto Taverna and Iv{\'a}n Agudo and Lucio Angelo Antonelli and Matteo Bachetti and Luca Baldini and Wayne H. Baumgartner and Ronaldo Bellazzini and Stefano Bianchi and Stephen D. Bongiorno and Raffaella Bonino and Alessandro Brez and Niccol{\'o} Bucciantini and Fiamma Capitanio and Simone Castellano and Elisabetta Cavazzuti and Chien-Ting J. Chen and Stefano Ciprini and Enrico Costa and Alessandra De Rosa and Ettore Del Monte and Laura Di Gesu and Niccol{\'o} Di Lalla and Immacolata Donnarumma and Michal Dovc̆iak and Steven R. Ehlert and Teruaki Enoto and Yuri Evangelista and Sergio Fabiani and Riccardo Ferrazzoli and Javier A. Garc{\'i}a and Shuichi Gunji and Kiyoshi Hayashida and Jeremy S. Heyl and Wataru Buz Iwakiri and Svetlana G. Jorstad and Philip Kaaret and Vladim{\'i}r Karas and Fabian Kislat and Takao Kitaguchi and Jeffery J. Kolodziejczak and Henric Krawczynski and Luca Latronico and Ioannis Liodakis and Simone Maldera and Alberto Manfreda and Fr{\'e}d{\'e}ric Marin and Andrea Marinucci and Alan P. Marscher and Francesco Massaro and Giorgio Matt and Ikuyuki Mitsuishi and Tsunefumi Mizuno and Fabio Muleri and Michela Negro and Chi-Yung Ng and Stephen L. O’Dell and Nicola Omodei and Chiara Oppedisano and Alessandro Papitto and George G. Pavlov and Abel Lawrence Peirson and Matteo Perri and Melissa Pesce-Rollins and Pierre-Olivier Petrucci and Maura Pilia and Simonetta Puccetti and Brian D. Ramsey and John Rankin and Ajay Ratheesh and Oliver J. Roberts and Roger W. Romani and Carmelo Sgro’ and Patrick O. Slane and Paolo Soffitta and Gloria Spandre and Douglas A. Swartz and Toru Tamagawa and Fabrizio Tavecchio and Yuzuru Tawara and Allyn F. Tennant and Nicholas E. Thomas and Francesco Tombesi and Alessio Trois and Roberto Turolla and Jacco Vink and Martin. C. Weisskopf and Kinwah Wu and Fei Xie and Silvia Zane},
  journal={Astronomy \& Astrophysics},
  year={2024},
  doi={10.1051/0004-6361/202450937}
}

@article{Mszros1988AstrophysicalIA,
  title={Astrophysical implications and observational prospects of X-ray polarimetry},
  author={P{\'e}ter M{\'e}sz{\'a}ros and Robert Novick and Andrew Szentgyorgyi and Gary A. Chanan and Martin. C. Weisskopf},
  journal={The Astrophysical Journal},
  year={1988},
  volume={324},
  pages={1056-1067},
  doi={10.1086/165962}
}

@article{Caiazzo2020PolarizationOA,
  title={Polarization of accreting X-ray pulsars. I. A new model},
  author={Ilaria Caiazzo and Jeremy S. Heyl},
  journal={Monthly Notices of the Royal Astronomical Society},
  year={2020},
  doi={10.1093/mnras/staa3428}
}

@article{Forsblom2025RevealingTO,
author = {{Forsblom}, Sofia V. and {Tsygankov}, Sergey S. and {Suleimanov}, Valery F. and {Mushtukov}, Alexander A. and {Poutanen}, Juri},
        title = "{Revealing two orthogonally polarized spectral components in Vela X-1 with IXPE}",
      journal = {\aap},
         year = 2025,
        month = apr,
       volume = {696},
          eid = {A224},
        pages = {A224},
          doi = {10.1051/0004-6361/202553867},
archivePrefix = {arXiv},
       eprint = {2501.14324},
 primaryClass = {astro-ph.HE},
       adsurl = {https://ui.adsabs.harvard.edu/abs/2025A&A...696A.224F}
}

@article{Soffitta2021TheIO,
  title={The Instrument of the Imaging X-Ray Polarimetry Explorer},
  author={Paolo Soffitta and Luca Baldini and Ronaldo Bellazzini and Enrico Costa and Luca Latronico and Fabio Muleri and Ettore Del Monte and Sergio Fabiani and M. Minuti and Michele Pinchera and Carmelo Sgro’ and Gloria Spandre and Alessio Trois and Fabrizio Amici and Hans E B Andersson and Primo Attin{\`a} and Matteo Bachetti and Mattia Barbanera and Fabio Borotto and Alessandro Brez and Daniele Brienza and Ciro Caporale and Claudia Cardelli and Rita Carpentiero and Simone Castellano and Marco Maria Castronuovo and Luca Cavalli and Elisabetta Cavazzuti and Marco Ceccanti and Mauro Centrone and Stefano Ciprini and Saverio Citraro and Fabio D’Amico and Elisa D’Alba and Sergio Di Cosimo and Niccolo’ Di Lalla and Alessandro Di Marco and Giuseppe Di Persio and Immacolata Donnarumma and Yuri Evangelista and Riccardo Ferrazzoli and Asami Hayato and Takao Kitaguchi and Fabio La Monaca and Carlo Lefevre and Pasqualino Loffredo and Paolo Lorenzi and Leonardo Lucchesi and C. Magazz{\`u} and Simone Maldera and Alberto Manfreda and Elio Mangraviti and Marco Marengo and Giorgio Matt and Paolo Mereu and Alfredo Morbidini and Federico Mosti and Toshio Nakano and Hikmat Nasimi and Barbara Negri and Seppo Arvo Anter Nenonen and Alessio Nuti and Leonardo Orsini and Matteo Perri and Melissa Pesce-Rollins and Raffaele Piazzolla and Maura Pilia and Alessandro Profeti and Simonetta Puccetti and John Rankin and Ajay Ratheesh and Alda Rubini and Francesco Santoli and Paolo F. Sarra and Emanuele Scalise and Andrea Sciortino and Toru Tamagawa and Marcello Tardiola and Antonino Tobia and Marco Vimercati and Fei Xie},
  journal={The Astronomical Journal},
  year={2021},
  volume={162},
  doi={10.3847/1538-3881/ac19b0}
}

@article{Weisskopf2022ImagingXP,
  title={Imaging X-ray Polarimetry Explorer: prelaunch},
  author={Martin. C. Weisskopf and Paolo Soffitta and Luca Baldini and Brian D. Ramsey and Stephen L. O’Dell and Roger W. Romani and Giorgio Matt and William D. Deininger and Wayne H. Baumgartner and Ronaldo Bellazzini and Enrico Costa and Jeffery J. Kolodziejczak and Luca Latronico and Herman L. Marshall and Fabio Muleri and Stephen D. Bongiorno and Allyn F. Tennant and Niccolo’ Bucciantini and Michal Dovc̆iak and Fr{\'e}d{\'e}ric Marin and Alan Marscher and Juri Poutanen and Patrick O. Slane and Roberto Turolla and William Kalinowski and Alessandro Di Marco and Sergio Fabiani and M. Minuti and Fabio La Monaca and Michele Pinchera and John Rankin and Carmelo Sgro’ and Alessio Trois and Fei Xie and Cheryl D. Alexander and David Allen and Fabrizio Amici and Jason J. Andersen and Angelo Antonelli and Spencer Antoniak and Primo Attin{\`a} and Mattia Barbanera and Matteo Bachetti and Randy M. Baggett and Jeff J. Bladt and Alessandro Brez and Raffaella Bonino and Christopher Boree and Fabio Borotto and Shawn Breeding and Daniele Brienza and Henry Kyle Bygott and Ciro Caporale and Claudia Cardelli and Rita Carpentiero and Simone Castellano and Marco Maria Castronuovo and Luca Cavalli and Elisabetta Cavazzuti and Marco Ceccanti and Mauro Centrone and Saverio Citraro and Fabio D’Amico and Elisa D’Alba and Laura Di Gesu and Ettore Del Monte and Kurtis L. Dietz and Niccolo’ Di Lalla and Giuseppe Di Persio and D. Dolan and Immacolata Donnarumma and Yuri Evangelista and Kevin Ferrant and Riccardo Ferrazzoli and MacKenzie Ferrie and Joseph N. Footdale and Brent Forsyth and Michelle Foster and Benjamin Garelick and Shuichi Gunji and Eli Gurnee and Michael Head and Grant Hibbard and Samantha Johnson and Erik M. Kelly and Kiranmayee Kilaru and Carlo Lefevre and Shelley Le Roy and Pasqualino Loffredo and Paolo Lorenzi and Leonardo Lucchesi and Tyler Maddox and Guido Magazz{\'u} and Simone Maldera and Alberto Manfreda and Elio Mangraviti and Marco Marengo and Alessandra Marrocchesi and Francesco Massaro and David T. Mauger and Jeffrey McCracken and Michael E. McEachen and Rondal Mize and Paolo Mereu and Scott Mitchell and Ikuyuki Mitsuishi and Alfredo Morbidini and Federico Mosti and Hikmat Nasimi and Barbara Negri and Michela Negro and Toan Nguyen and Isaac Nitschke and Alessio Nuti and Mitch Onizuka and Chiara Oppedisano and Leonardo Orsini and Darren Osborne and Richard Pacheco and Alessandro Paggi and William Painter and Steven D. Pavelitz and Christina Pentz and Raffaele Piazzolla and Matteo Perri and Melissa Pesce-Rollins and Colin A. Peterson and Maura Pilia and Alessandro Profeti and Simonetta Puccetti and Jaganathan Ranganathan and Ajay Ratheesh and Leroy H. Reedy and Noah Root and Alda Rubini and Stephanie Ruswick and Javier Sanchez and Paolo F. Sarra and Francesco Santoli and Emanuele Scalise and Andrea Sciortino and Christopher Schroeder and Timothy Seek and Kalie Sosdian and Gloria Spandre and Chet O. Speegle and Toru Tamagawa and Marcello Tardiola and Antonino Tobia and Nicholas E. Thomas and Robert Valerie and Marco Vimercati and Amy L. Walden and Bruce Weddendorf and Jeffrey Wedmore and David Welch and Davide Zanetti and Francesco Zanetti},
  journal={Journal of Astronomical Telescopes, Instruments, and Systems},
  year={2022},
  volume={8},
  pages={026002 - 026002},
  doi={10.1117/1.JATIS.8.2.026002}
}

@article{Baldini2022ixpeobssimAS,
  title={ixpeobssim: A simulation and analysis framework for the imaging X-ray polarimetry explorer},
  author={L. Baldini and Niccol{\'o} Bucciantini and N. Di Lalla and Steven R. Ehlert and Alberto Manfreda and Michela Negro and Nicola Omodei and Melissa Pesce-Rollins and Carmelo Sgro’ and Stefano Silvestri},
  journal={SoftwareX},
  year={2022},
  volume={19},
  pages={101194},
  doi={10.1016/j.softx.2022.101194}
}

@article{Grinberg2017TheCA,
  title={The clumpy absorber in the high-mass X-ray binary Vela X-1},
  author={Victoria Grinberg and Natalie Hell and Ileyk El Mellah and Joey Neilsen and Andreas A. C. Sander and Maurice A. Leutenegger and Maurice A. Leutenegger and Felix S. F{\"u}rst and David P. Huenemoerder and Peter Kretschmar and Matthias Bissinger n{\'e} K{\"u}hnel and Silvia Mart{\'i}nez-N{\'u}{\~n}ez and Shu Niu and Shu Niu and Katja Pottschmidt and Katja Pottschmidt and Norbert S. Schulz and J{\"o}rn Wilms and Michael A. Nowak},
  journal={arXiv: High Energy Astrophysical Phenomena},
  year={2017},
  doi={10.1051/0004-6361/201731843}
}

@article{Diez2022ContinuumCL,
  title={Continuum, cyclotron line, and absorption variability in the high-mass X-ray binary Vela X-1},
  author={Cristina Montero Diez and Victoria Grinberg and F. Furst and E. Sokolova-Lapa and Andrea Santangelo and J{\"o}rn Wilms and Katja Pottschmidt and S. Mart'inez-N'unez and Christian Malacaria and Peter Kretschmar},
  journal={Astronomy \& Astrophysics},
  year={2022},
  doi={10.1051/0004-6361/202141751}
}

@article{Diez2023ObservingTO,
  title={Observing the onset of the accretion wake in Vela X-1},
  author={Cristina Montero Diez and Victoria Grinberg and F. Furst and Ileyk El Mellah and M. Zhou and Andrea Santangelo and S. Mart'inez-N'unez and Roberta Amato and Natalie Hell and Peter Kretschmar},
  journal={Astronomy \& Astrophysics},
  year={2023},
  doi={10.1051/0004-6361/202245708}
}

@article{MartinezNunez2014TheAE,
  title={The accretion environment in Vela X-1 during a flaring period using XMM-Newton},
  author={Silvia Mart'inez-N'unez and Jos'e Miguel Torrej'on and Matthias Kuhnel and Peter Kretschmar and Martin Stuhlinger and Jose Joaqu{\'i}n Rodes-Roca and F. Furst and Ingo Kreykenbohm and Antonio Martin-Carrillo and Andrew M. T. Pollock and Joern Wilms},
  journal={Astronomy and Astrophysics},
  year={2014},
  volume={563},
  doi={10.1051/0004-6361/201322404}
}

@article{Poutanen2024XrayPO,
  title={X-ray Polarimetry of X-ray Pulsars},
  author={Juri Poutanen and Sergey S. Tsygankov and Sofia V. Forsblom},
  journal={Galaxies},
  year={2024},
  doi={10.3390/galaxies12040046}
}

@article{Kallman2015XRAYPF,
  title={X-RAY POLARIZATION FROM HIGH-MASS X-RAY BINARIES},
  author={Timothy R. Kallman and Anton V. Dorodnitsyn and John M. Blondin},
  journal={The Astrophysical Journal},
  year={2015},
  volume={815},
  doi={10.1088/0004-637X/815/1/53}
}

@article{Fuerst2013NuSTARDO,
  title={NuSTAR DISCOVERY OF A LUMINOSITY DEPENDENT CYCLOTRON LINE ENERGY IN VELA X-1},
  author={Felix Fuerst and Katja Pottschmidt and Joern Wilms and John A. Tomsick and Matteo Bachetti and Steven E. Boggs and Finn E. Christensen and William W. Craig and Brian W. Grefenstette and Charles J. Hailey and Fiona A. Harrison and Kristin Kruse Madsen and Jonathan M. Miller and Daniel K. Stern and Dominic J. Walton and Will Zhang},
  journal={The Astrophysical Journal},
  year={2013},
  volume={780},
  doi={10.1088/0004-637X/780/2/133}
}

@article{Rahin2023ANV,
  title={A NICER Viewing Angle on the Accretion Stream of Vela X-1},
  author={Roi Rahin and Ehud Behar},
  journal={The Astrophysical Journal},
  year={2023},
  volume={950},
  doi={10.3847/1538-4357/acc386}
}

@article{ForemanMackey2012emceeTM,
  title={emcee: The MCMC Hammer},
  author={Daniel Foreman-Mackey and David W. Hogg and Dustin Lang and Jonathan B. Goodman},
  journal={Publications of the Astronomical Society of the Pacific},
  year={2012},
  volume={125},
  pages={306 - 312},
  doi={D10.1086/670067}
}

@article{Hickox2024,
  title={Origin of the Soft Excess in X-Ray Pulsars},
  author={Ryan C. Hickox and Ramesh Narayan and Timothy R. Kallman},
  journal={The Astrophysical Journal},
  year={2004},
  volume={614},
  pages={881 - 896},
  doi={10.1086/423928}
}

@article{DiMarco2023HandlingTB,
  title={Handling the Background in IXPE Polarimetric Data},
  author={Alessandro {Di Marco} and Paolo Soffitta and Enrico Costa and Riccardo Ferrazzoli and Fabio {La Monaca} and John Rankin and Ajay Ratheesh and Fei Xie and Luca Baldini and Ettore {Del Monte} and Steven R. Ehlert and Sergio Fabiani and Dawoon E. Kim and Fabio Muleri and Stephen L. O’Dell and Brian D. Ramsey and Alda Rubini and Carmelo Sgro’ and Stefano Silvestri and Allyn F. Tennant and Martin. C. Weisskopf},
  journal={The Astronomical Journal},
  year={2023},
  volume={165},
  doi={10.3847/1538-3881/acba0f}
}

@ARTICLE{Arnaud1996,
       author = {{Arnaud}, K.~A.},
        title = "{XSPEC: The First Ten Years}",
    booktitle = {Astronomical Data Analysis Software and Systems V},
         year = 1996,
       editor = {{Jacoby}, George H. and {Barnes}, Jeannette},
       series = {Astronomical Society of the Pacific Conference Series},
       volume = {101},
        month = jan,
        pages = {17},
       adsurl = {https://ui.adsabs.harvard.edu/abs/1996ASPC..101...17A}
}

@article{Garg2023,
  title={Flux-resolved Spectropolarimetric Evolution of the X-Ray Pulsar Hercules X-1 Using IXPE},
  author={Garg Akash and Rawat Divya and Bhargava Yash sand Méndez Mariano and Bhattacharyya Sudip},
  journal={The Astrophysical Journal Letters},
  year={2023},
  volume={948},
  doi={10.3847/2041-8213/acccfa}
}

@article{Gnedin1978TheEO,
       author = {{Gnedin}, Iu. N. and {Pavlov}, G.~G. and {Shibanov}, Iu. A.},
        title = "{The effect of vacuum birefringence in a magnetic field on polarization and directivity of radiation of X-ray pulsars.}",
      journal = {Pisma v Astronomicheskii Zhurnal},
         year = 1978,
        month = jun,
       volume = {4},
        pages = {214-218},
       adsurl = {https://ui.adsabs.harvard.edu/abs/1978PAZh....4..214G}
}

@article{Pavlov1979InfluenceOV,
  title={Influence of vacuum polarization by a magnetic field on the propagation of electromagnetic waves in a plasma},
  author={George G. Pavlov and Y. A. Shibanov},
  journal={Journal of Experimental and Theoretical Physics},
  year={1979},
  volume={49},
  pages={1457-1473}
}

@article{Gnedin1974,
  title={The transfer equations for normal waves and radiation polarization in an anisotropic medium},
  author={Gnedin, Y.N.},
  journal={Sov. J. Exp. Theor. Phys.},
  year={1974},
  volume={38},
  pages={903-908}
}
\bibliographystyle{aasjournal}

\end{document}